\documentclass[11pt]{article}
\usepackage{amsmath}
\usepackage{amsfonts}
\usepackage{graphicx}
\usepackage{cancel}
\usepackage{fullpage}
\usepackage{hyperref}
\usepackage{color}

\newcommand{\order}[1]{\mathcal{O}\!\left(#1\right)}

\newcommand{\derOrd}[2]{\frac{d#1}{d#2}}
\newcommand{\der}[2]{\frac{\partial#1}{\partial#2}}

\newcommand{\non}{\nonumber\\}

\numberwithin{equation}{section}

\title{Fokker-Planck approach to wave turbulence}
\date{}
\author{Daniel Schubring}
\begin{document}
	
	\maketitle
	
	\begin{center}
\it{Initiative for the Theoretical Sciences}\\ \it{ The Graduate Center, CUNY}\\ \it{
	365 Fifth Ave, New York, NY 10016, USA}\\[.5cm]
	\end{center}

		The Kolmogorov-Zakharov stationary states for weak wave turbulence involve solving a leading-order kinetic equation. Recent calculations of higher-order corrections to this kinetic equation using the Martin-Siggia-Rose path integral are reconsidered in terms of stationary states of a Fokker-Planck Hamiltonian. A non-perturbative relation closely related to the quantum mechanical Ehrenfest theorem is introduced and used to express the kinetic equation in terms of divergences of two-point expectation values in the limit of zero dissipation. Similar equations are associated to divergences in higher-order cumulants. It is additionally shown that the ordinary thermal equilibrium state is not actually a stationary state of the Fokker-Planck Hamiltonian, and a non-linear modification of dissipation is considered to remedy this.
	\tableofcontents

	\section{Introduction}\label{Sec intro}
	A fundamental result in the theory of weak wave turbulence was the discovery of the Kolmogorov-Zakharov (KZ) solutions \cite{Zakharov}, which are stationary states analogous to the Kolmogorov spectrum in hydrodynamic turbulence, and which are distinct from the thermal equilibrium (Rayleigh-Jeans) solution. The KZ solutions have considerable numerical and experimental support, see e.g. \cite{NewellRumpf2011,ZLF.text,Nazarenko.text} for reviews. These are solutions to a kinetic equation which is derived from low-order perturbation theory \cite{Hasselmann1962}. The non-perturbative existence of the KZ solutions and even their higher-order corrections are still open theoretical questions.
	
	Recently there has been renewed progress in the calculation of higher-order corrections to the kinetic equation  \cite{RS.1,RSSS, PavezDuring2023} using a stochastic Martin-Siggia-Rose (MSR) path integral approach. The MSR path integral \cite{MSR,DeDominicis1975,Janssen1976} has been used extensively in hydrodynamic turbulence \cite{Eyink1996,AdzhemyanAntonovVasiliev1999}, but usually not as explicitly within wave turbulence, although it is equivalent \cite{BereraSalewskiMcComb2013} to a suitably corrected Wyld method \cite{Wyld1961,Lee1965}. Usually these approaches are treated perturbatively, but this paper intends to provide a theoretical framework for discussing the non-perturbative KZ stationary state, and it makes the connection explicit between the MSR methods and the phase space methods of Gurarie \cite{Gurarie.1} and Rosenhaus, Smolkin \cite{RS.2}.
	
	In both \cite{Gurarie.1} and \cite{RS.2}, the KZ solution is treated as a probability distribution  $\rho$ on phase space which is stationary in time. Time-independent distributions obey the Liouville equation,
	\begin{align}
		\der{\rho}{t}=-\{\rho,H\}=0. \label{Liouville}
	\end{align}
One physically motivated solution is the thermal equilibrium (``Boltzmann'') solution
\begin{align}
	\rho_{B}=e^{-\frac{H}{T}},\label{Def rhoB}
\end{align}
but clearly any general function of $H$ alone will also be time-independent. We will later modify the Liouville equation to incorporate random forcing and dissipation, and then the special form of $\rho_B$ will be singled out. For now let us continue to discuss the Hamiltonian dynamics alone.

A stationary distribution may also depend on other conserved quantities. In the special case that the Hamiltonian $H_0$ is just a non-interacting collection of harmonic oscillators\footnote{To be clear, the frequencies $\omega_k$ and temperatures $T_k$ have dimensions of energy and $q_k$ and $p_k$ are dimensionless canonical coordinates related to the usual harmonic oscillator position and momentum by a trivial rescaling.} indexed by the mode number $k$
\begin{align}
	H_0=\sum_k \frac{\omega_k}{2} \left(p_k^2 + q_k^2\right),
\end{align}
the wave action $J_k = \frac{1}{2}\left(p_k^2 + q_k^2\right)$ for each individual mode is an independent conserved quantity, and so there are also `non-equilibrium' stationary states $\rho_0$ with an independent temperature $T_k$ for each mode,
\begin{align}
	\rho_0 = e^{-\sum_k \frac{\omega_k}{T_k}J_k}.\label{Def rho0}
\end{align}
 Once again, a more general non-Gaussian function of the $J_k$ will also satisfy the stationary Liouville equation, and we will discuss this possibility later. For now, note that even given the special form of $\rho_0$ in \eqref{Def rho0}, there is a lot of freedom in the choice of the $T_k$, and this allows us to arbitrarily fix the wave action spectrum $n_k$,
 \begin{align}
 	n_k \equiv \langle J_k\rangle = \frac{T_k}{\omega_k}.
 \end{align}
 To single out the KZ spectrum $n_k$, we need to introduce an interacting Hamiltonian $H= H_0+V$. To any given order in the perturbation $V$ we may solve for the distribution $\rho$ that satisfies the stationary Liouville equation \eqref{Liouville} and which reduces to $\rho_0$ at lowest order. It will turn out that this construction is pathological unless the $T_k$ take a special form, and this is what defines the KZ state.
 
 In \cite{Gurarie.1} the corrections to $\rho_0$ are found by finding the corrections to $\sum_k\omega_k J_k/T_k$ such that the quantity is conserved under the flow of $H$ to any given order in perturbation theory. This is not possible for generic $T_k$ due to resonances (the old problem of divergent ``small denominators''), and a regulating $i\epsilon$ is introduced by hand to deal with this.

A similar approach is taken in \cite{RS.2}, but instead the corrections to $\rho_0$ are found through the analogue of the Lippmann-Schwinger equation for a Hamiltonian $\hat{H}_L$ associated to the Liouville equation, as in Prigogine's work \cite{Prigogine}. Once again divergences are encountered, and an $i\epsilon$ term is introduced. In this paper the Liouville Hamiltonian $\hat{H}_L$ will be extended to a more well-behaved Fokker-Planck Hamiltonian and the $i\epsilon$ regulator will arise naturally due to dissipation.

So in principle there is a method to calculate $\rho$ to any finite order in perturbation theory, and expectation values of some function $O(q,p)$ on phase space may be calculated with an integral
\begin{align}
\langle O\rangle = \frac{1}{Z}\int \mathcal{D}q\mathcal{D}p\,O(q,p)\rho(q,p).
\end{align}
This is just a generalization of the equilibrium partition function where an arbitrary non-equilibrium stationary state $\rho$ is used in place of the thermal equilibrium $\rho_B$. Path integral notation $\mathcal{D}q\mathcal{D}p$ is used for the integration variables $\prod_k dq_k dp_k$ since in practice we will be dealing with classical field theories where $k$ is treated as a continuous argument. When emphasizing the distinction is necessary, this integral over phase space will be referred to as the \emph{stationary state path integral}.

 In \cite{RS.1} a distinct path integral is introduced following the MSR formalism, and which will be referred to as the \emph{stochastic path integral}. This approach explicitly involves introducing random forcing and dissipation unlike the stationary state approaches \cite{Gurarie.1,RS.2} discussed above. The stochastic path integral is quite different from the stationary state path integral since the fields involved are time dependent. The stochastic path integral is potentially more powerful since correlation functions at unequal times may be calculated, but for the purposes of calculating corrections to the kinetic equation only equal time expectation values will be needed. In \cite{RSSS} we derived simple rules for calculating equal time expectation values in the MSR approach, and the result was shown to agree with the stationary state approaches in the limit of vanishing forcing and dissipation.

In this paper the relation between the stationary state approaches and the stochastic path integral approach is clarified. This is done by considering the \emph{Fokker-Planck Hamiltonian} which has the same relation to the stochastic path integral as the quantum mechanical Hamiltonian has to the Euclidean path integral in quantum mechanics. The Fokker-Planck Hamiltonian $\hat{H}$ is just the sum of the Liouville Hamiltonian $\hat{H}_L$ appearing in \cite{RS.2} (referred to as $iL$ there) and a part that depends on dissipation $\hat{H}_\gamma$ and which helps regulate the theory.

The Fokker-Planck approach considered here helps clarify certain non-perturbative features of the stochastic path integral approach. It is well-known that the thermal equilibrium case with all temperatures $T_k$ equal to some constant $T$ is also a solution to the leading-order kinetic equation, and in \cite{RS.1} the authors also test the subleading corrections to the kinetic equation by considering the thermal equilibrium special case. In the limit of vanishing dissipation the thermal equilibrium state $\rho_B$ is always a solution of the Liouville equation so this makes sense.
%\footnote{This is usually written in terms of occupation numbers $n_k$ defined in \eqref{Def n}. Thermal equilibrium corresponds to setting $n_k=T/\omega_k$.}

However there are many other stationary solutions to the Liouville equation besides $\rho_B$, and in order to single out a unique stationary state (at least in this approach) we need to introduce some small random forcing and dissipation. It is shown that with the linear form of dissipation in \cite{RS.1,RSSS} $\rho_B$ is not actually a stationary state of the Fokker-Planck Hamiltonian. A modified \emph{non-linear dissipation} term which has some physical motivation \cite{SPDas.text} is introduced to correct this in Sec \ref{Sec non-linear diss intro}.

Both the conventional linear dissipation and non-linear dissipation will be used throughout this paper, and for many purposes calculations with the linear dissipation choice are simpler. But the non-linear dissipation term has one practical advantage for theoretical calculations. Upon setting all of the temperatures $T_k=T$ in the non-linear dissipation case, the exact stationary state becomes $\rho_B$. Since $\rho_B$ has no dependence on dissipation at all  expectation values will simplify dramatically, even without taking a vanishing dissipation limit. Since the expressions for higher-order corrections to expectation values can be very complicated this is a strong consistency check on calculations. This is discussed further in Sec \ref{Sec non-linear diss eq}, after first discussing some details of the stochastic path integral for non-linear dissipation in Sec \ref{Sec non-linear diss jacobian}.

Another non-perturbative question clarified by the Fokker-Planck approach has to do with how the classical equations of motion are manifested in expectation values. Since the action in the stochastic path integral involves the equations of motion squared rather than the classical Lagrangian, it is not as straightforward as using the Ward identities familiar from quantum field theory. Instead expectations of classical equations of motion will involve corrections due to dissipation and forcing.

This will be discussed further in Sec \ref{Sec ehrenfest}, where the correct relation is shown to be somewhat similar to the Ehrenfest theorem in ordinary quantum mechanics. Using this \emph{stochastic Ehrenfest theorem}, in Sec \ref{Sec kinetic eq} the vanishing of the collision integral in the kinetic equation is shown to be equivalent to regularity of two-point expectation values in the limit of zero dissipation. Higher-order cumulants may also be considered in this manner, and in  Sec \ref{Sec ehrenfest JJ} it is shown that they seem to lead to equations independent of the usual kinetic equation.

Ultimately the Fokker-Planck approach gives results equivalent to the stochastic path integral approach, and this is shown further in Sec \ref{Sec corrections} and the associated Appendix \ref{Sec additional calculations} where the corrections to the stationary state are calculated explicitly. At finite dissipation no divergences are encountered and the expectation values thus calculated are shown to agree with \cite{RS.1} and \cite{RSSS}. 

Below in Sec \ref{Sec FP Ham} we begin by introducing the equations of motion together with forcing and dissipation and deriving the Fokker-Planck Hamiltonian. Sec \ref{Sec discussion} continues the general discussion about this stochastic approach which involves auxiliary forcing and dissipation even in the inertial range, and relates it to a more usual time-dependent approach in the zero dissipation limit.

\section{The Fokker-Planck Hamiltonian}\label{Sec FP Ham}
In the general case, consider a set of Langevin equations of the form
\begin{align}
	\dot{x}_a(t)=V_a(\mathbf{x}(t))+f_a(t),
\end{align}
where $f_a$ is a stochastic forcing term, with correlations
\begin{align}
	\left\langle f_a(t_1)f_b(t_2)\right\rangle = F_a\delta_{ab}\delta(t_1-t_2).\label{forcing function}
\end{align}
We will be more specific below, but in our context the real quantities $x_a$ represent coordinates on phase space, and the function $V_a$ encodes both Hamilton's equations of motion and a dissipation term.
 
Following standard arguments,\footnote{For a brief derivation see section 34.2 of \cite{ZJ.text}. The variables $x, V, f, F, \rho$ in our notation are respectively denoted $q, -\frac{1}{2}f, \nu, \Omega, P$ there.} the Langevin equations imply a Fokker-Planck equation for the evolution of a probability distribution $\rho$ over $x$,
\begin{gather}
	\der{}{t}\rho\left(\mathbf{x}, t\right)=-\hat{H}\rho\left(\mathbf{x}, t\right),\label{FP eq general}\\
	\hat{H}\rho=\sum_a\der{}{x_a}\left(V_a\rho-\frac{F_a}{2}\der{\rho}{x_a}\right).\label{FP H general}
\end{gather}
The operator $\hat{H}$ acting on functions over $x$ is the \emph{Fokker-Planck Hamiltonian}. It should not be confused with the canonical Hamiltonian $H$ which is a function on phase space, and hat notation will be used to keep the two notions distinct.

We will apply this general expression to a classical field theory with forcing and dissipation. In our context $x_a$ will represent either a field $q_k$ or its conjugate momentum $p_k$, both indexed by a Fourier mode index $k$ that we will treat as discrete throughout. The equations of motion are
\begin{align}
	\dot{q}_k(t)&=\der{H}{p_k}-\frac{\gamma_k}{\omega_k} \der{H_0}{q_k}+f_{q,k},\non
	\dot{p}_k(t)&=-\der{H}{q_k}-\frac{\gamma_k}{\omega_k} \der{H_0}{p_k}+f_{p,k}.\label{eq motion real parts}
\end{align}
Since the unperturbed Hamiltonian $H_0$ is quadratic in $q_k$ and $p_k$ the dissipation terms are linear. This implementation of dissipation will be referred to as \emph{linear dissipation}.

These equations may be written in terms of a single complex variable per mode $a_k$
\begin{gather}
	a_k\equiv \frac{q_k+ip_k}{\sqrt{2}},\\
%	\partial_k\equiv \frac{1}{\sqrt{2}}\left(\der{}{\varphi_k}-i\der{}{\pi_k}\right),\qquad 	\bar{\partial}_k\equiv \frac{1}{\sqrt{2}}\left(\der{}{\varphi_k}+i\der{}{\pi_k}\right),\\
		\dot{a}_k=-i\bar{\partial}_k{H}-\gamma_k a_k + f_k,\qquad 	f_k\equiv \frac{f_{q,k}+if_{p,k}}{\sqrt{2}}.\label{eq motion a}
\end{gather}
Here bars are used to indicate the complex conjugate and the notation $\partial_k, \bar{\partial}_k$ is used for derivatives with respect to $a_k$ and $\bar{a}_k$.

Applying the general expression \eqref{FP H general} to the equations \eqref{eq motion real parts}, the Fokker-Planck Hamiltonian breaks up into a \emph{Liouville Hamiltonian} $\hat{H}_L$ and a \emph{dissipative Hamiltonian} $\hat{H}_\gamma$,
\begin{gather}
\hat{H}=\hat{H}_L+\hat{H}_\gamma,\\
\hat{H}_L\rho = \{\rho, H\}=-i\sum_k \left(\partial_k\rho \bar{\partial}_k H-\partial_k H \bar{\partial}_k\rho\right),\\
\hat{H}_\gamma\rho=-\sum_k\frac{\gamma_k}{\omega_k}\bar{\partial}_{k}\left[\left(\partial_k H_0\right)\rho +T_k\partial_k \rho\right]+\text{c.c.},\qquad\text{(linear dissipation)}\label{Hgamma linear}
\end{gather}
where the quantity $T_k$ is related to the strength of the forcing function \eqref{forcing function} through $F_k = 2\frac{\gamma_k}{\omega_k}T_k$.

In the free case $H=H_0$ there is an obvious stationary distribution $\hat{H}_0 \rho_0 = 0$ given by \eqref{Def rho0},
\begin{align*}
	\rho_0 = e^{-\sum_k \frac{\omega_k}{T_k} \bar{a}_k a_k},
\end{align*}
so $T_k$ may be interpreted as a distinct temperature for each mode $k$. It is also directly related to the wave action spectrum,
\begin{align}
	n_k = \langle \bar{a}_k a_k\rangle^{(0)} = \frac{T_k}{\omega_k}.\label{Def n}
\end{align}
Here the superscript $(0)$ indicates the average is taken with respect to $\rho_0$, and more generally superscripts will indicate the order of an expectation value in perturbation theory.

To discuss higher order corrections $\rho=\rho_0+\rho_1+\dots$ concretely, we must be clear about the form of our perturbation  $V$. There are two simple examples of interactions which we will treat here and which are commonly considered in wave turbulence. The cubic or \emph{three-wave} case,
\begin{align}
	V=\frac{1}{2}\sum_{kij}\left(  \lambda_{k;ij}\bar{a}_k a_i a_j + \bar{\lambda}_{k;ij}a_k \bar{a}_i \bar{a}_j\right),\label{V three-wave}
\end{align}
and the quartic or \emph{four-wave} case,
\begin{align}
	V=\sum_{ijkl}\lambda_{ij;kl}\bar{a}_i \bar{a}_j a_k a_l,\qquad \left(\bar{\lambda}_{ij;kl} = \lambda_{kl;ij}\right).\label{V four-wave}
\end{align}
In physical applications the mode indices refer to the Fourier transform of the position coordinate of fields, and are usually taken to be continuous. In particular this is essential for the KZ spectra solutions. But since we will not explicitly carry out integrals over the modes in this paper, we will treat them as discrete for notational convenience. The perturbative parameters $\lambda_{k;ij}$ and $\lambda_{ij;kl}$ themselves depend on mode indices, and they are taken to contain a delta function such that $\lambda_{k;ij}=0$ unless the momentum associated to $i$ plus $j$ equals that associated to $k$, and likewise for the four-wave case.

\subsection{Introducing non-linear dissipation}\label{Sec non-linear diss intro}
When all modes are driven at the same temperature $T_k=T$, one might expect the thermal equilibrium state $\rho_{B}=e^{-\frac{H}{T}}$ to be a stationary state of the interacting Fokker-Planck equation, where $H=H_0+V$. It is a stationary state of the Liouville Hamiltonian $\hat{H}_L$, but $\hat{H}_\gamma\rho_{B}\neq 0$, so the thermal equilibrium state is not stationary.

This suggests a simple modification of the dissipation terms in \eqref{eq motion real parts} where the full Hamiltonian $H$ appears rather than $H_0$. In complex form, the equations of motion are
\begin{align}
\dot{a}_k=-\left(i+\frac{\gamma_k}{\omega_k}\right)\bar{\partial}_k{H}+ f_k.\label{eq motion a nonpert}
\end{align}
This will be referred to as \emph{non-linear dissipation}. The only modification to the Fokker-Planck Hamiltonian is that the full $H$ appears in the dissipative part,
\begin{align}
	\hat{H}_\gamma\rho=-\sum_k\frac{\gamma_k}{\omega_k}\bar{\partial}_{k}\left[\left(\partial_k H\right)\rho +T_k\partial_k \rho\right]+\text{c.c.},\qquad\text{(non-linear dissipation)}\label{Hgamma nonperturbative}
\end{align}
and thus $\rho_{B}=e^{-\frac{H}{T}}$ is indeed a stationary state in the case where $T_k=T$.

Non-linear dissipation has some physical motivation. In the absence of forcing, it ensures that the fixed points of the dynamics are at the local minima of the full Hamiltonian $H$, whereas the linear dissipation term tends to drive the dynamics to the local minima of the unperturbed Hamiltonian $H_0$, which are generically different.\footnote{For the simple three-wave and four-wave perturbations that we will consider here both $H_0$ and $H$ have a local minimum at $a_k=0$, although if we modify the three-wave system by linear counterterms as in \eqref{tadpole counterterm} this is no longer true.} Perhaps more importantly, this form of dissipation may plausibly arise due to microscopic modes in a Hamiltonian system. See for instance Sec 6.1.2 of \cite{SPDas.text} where a very similar Langevin equation is derived for describing the slow modes of a dense fluid. In this more general case the dissipation is not only non-linear, but the dissipation coefficients (referred to as $L^0_{ij}$ in this context) may in principle be non-diagonal and depend on the Langevin equation variables themselves.

\section{Stochastic Ehrenfest theorem}\label{Sec ehrenfest}
The Fokker-Planck equation leads directly to relations between expectation values which are analogous to the Ehrenfest theorem in quantum mechanics.

Let $G$ be an arbitrary time-independent function on phase space, and define its expectation value with respect to some probability distribution $\rho$ by
\begin{align}
	\langle G\rangle(t) \equiv \frac{1}{Z}\int \mathcal{D}a\mathcal{D}\bar{a}\, G(a,\bar{a}) \rho(a,\bar{a},t),
\end{align}
Here $\int \mathcal{D}a\mathcal{D}\bar{a}$ denotes a phase space integral, and $Z\equiv \int \mathcal{D}a\mathcal{D}\bar{a}\rho$ is a normalization factor that is constant with time. Now take a time derivative using \eqref{FP eq general}, and integrate by parts in $\hat{H}_L$,
\begin{gather}
	\derOrd{}{t}\langle G\rangle= \left\langle\{G,H\}\right\rangle - \frac{1}{Z}\int \mathcal{D}a\mathcal{D}\bar{a}\,G\,\hat{H}_\gamma\rho.%\non
	%\left\langle\dot{G}\right\rangle = \frac{1}{Z}\int \mathcal{D}a\mathcal{D}\bar{a} G(a,\bar{a})\hat{H}_\gamma\rho
\end{gather}
This is in the form of the Ehrenfest theorem with an additional term involving $\hat{H}_\gamma$ that represents the correction due to dissipation and forcing.

Let us now specialize to the stationary state, $\hat{H}\rho=0$, so all expectation values are time independent. Using the expressions \eqref{Hgamma linear}\eqref{Hgamma nonperturbative} for $\hat{H}_\gamma$ and integrating by parts,
\begin{align}
	\left\langle\{G,H\}\right\rangle&= \sum_k \frac{\gamma_k}{\omega_k}\left\langle\bar{\partial}_k G \partial_k \left(H_0+\eta V\right) + {\partial}_k G \bar{\partial}_k \left(H_0+\eta V\right)\right\rangle-\frac{2\gamma_k T_k}{\omega_k}\left\langle\bar{\partial}_k \partial_k G\right\rangle.\label{ehrenfest}
\end{align}
In order to cover both forms of dissipation we have introduced the parameter $\eta$. We set $\eta=0$ for linear dissipation and $\eta=1$ for non-linear dissipation.

The interpretation of \eqref{ehrenfest} is clarified by taking a time derivative of $G$ and using the Langevin equation \eqref{eq motion a},
\begin{align}
		\derOrd{G}{t}&=\sum_k \partial_k G 	\derOrd{a_k}{t} + \bar{\partial}_k G \derOrd{\bar{a}_k}{t}\non
	&= \{G,H\} -\sum_k \frac{\gamma_k}{\omega_k}\left({\partial}_k G \bar{\partial}_k H_0+\bar{\partial}_k G \partial_k H_0 \right) +\left(\partial_k G f_k + \bar{\partial}_k G \bar{f}_k\right).\nonumber
\end{align}
Clearly the first term on the right-hand side of \eqref{ehrenfest} is just the effect of the dissipation term in Langevin equation. The second term represents the effect of the random forcing term, if we have the following rule for correlations between phase space functions $G'$ and $f_k$,
\begin{align}
	\langle G' f_k\rangle = \frac{1}{2}F_k \langle \bar{\partial}_k G'\rangle,\qquad \langle G' \bar{f}_k\rangle = \frac{1}{2}F_k \langle {\partial}_k G'\rangle,
\end{align}
where recall that $F_k$ is the strength of the forcing \eqref{forcing function}.

To make use of the stochastic Ehrenfest theorem \eqref{ehrenfest}, we will substitute some simple expressions for $G$. The very simplest case of linear $G=a_k$ leads to a non-perturbative result on the expectation value $\langle a_k\rangle$, and this is discussed in Appendix \ref{Sec non-perturbative tadpole}. A more important case arises upon choosing $G$ to be the equal time two-point function $\bar{a}_r a_r$.

\subsection{The collision integral}\label{Sec kinetic eq}
Choosing $G=\bar{a}_r a_r$ in \eqref{ehrenfest},
\begin{align}
	\left\langle\{\bar{a}_r a_r, H\}\right\rangle=2\gamma_r\left(\left\langle \bar{a}_r a_r\right\rangle-n_r\right)+\eta\frac{\gamma_r}{\omega_r}\left\langle a_r \partial_r V + \bar{a}_r\bar{\partial}_r V\right\rangle.\label{kinetic eq two sided}
\end{align}
The left-hand side is the \emph{collision integral} that appears in the time dependent wave kinetic equation $\derOrd{}{t}\langle \bar{a}_r a_r\rangle= \left\langle\{\bar{a}_r a_r,H\}\right\rangle$ in the $\gamma \rightarrow 0$ limit. In particular, for the three-wave and four-wave theories,
\begin{align}
	\left\langle\{\bar{a}_r a_r, H\}\right\rangle &= \sum_{ij}\text{Im}\left[\lambda_{r;ij}\langle \bar{a}_r a_i a_j\rangle-2\lambda_{i;jr}\langle \bar{a}_i a_j a_r\rangle\right]\qquad \text{(three-wave)}\label{kinetic eq three wave}\\ &= 4\sum_{jkl}\text{Im}\left[\lambda_{rj;kl}\langle \bar{a}_r \bar{a}_j a_k a_l\rangle\right].\qquad \text{(four-wave)}\label{kinetic eq four wave}
\end{align}
The explicit expressions for the lowest order connected three- and four-point expectation values are well-known, and are calculated in the approach of this paper in \eqref{three point exp} and \eqref{four point exp}, respectively.

\begin{figure}
	\centering
	\includegraphics[width=0.6\textwidth]{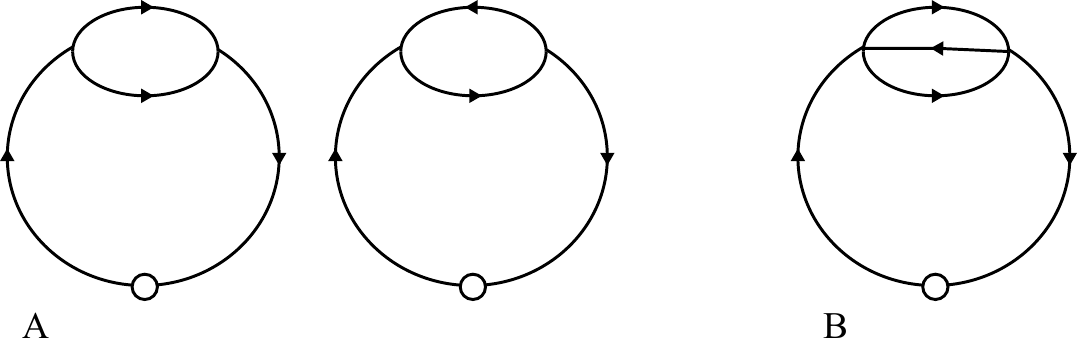}
	\caption{Diagrams for $\langle \bar{a} a\rangle^{(2)}$ which correspond to the lowest order collision integral. An open circle indicates the local in time operator $\bar{a} a$, and the diagrams indicate self-energy corrections. The two diagrams in (A) correspond to the two terms in the three-wave collision integral \eqref{kinetic eq three wave} and (B) corresponds to the four-wave collision integral \eqref{kinetic eq four wave}.}
	\label{figKinetic}
\end{figure}

For any stationary state that survives the limit of vanishing dissipation, the collision integral should vanish,
\begin{align}
	\lim_{\gamma\rightarrow 0} \left\langle\{\bar{a}_r a_r, H\}\right\rangle =0.\label{kinetic eq general}
\end{align}
This condition is non-trivial and it can not be satisfied for most choices of $T_k$ or equivalently $n_k$. One solution will be the thermal equilibrium solution where $T_k=T$, but there may also be non-equilibrium KZ solutions.

The right-hand side of \eqref{kinetic eq two sided} shows that the collision integral may be equivalently calculated in terms of the part of $\langle \bar{a}_r a_r\rangle$ that is proportional to $\gamma_r^{-1}$. Or put another way, the vanishing of the collision integral is equivalent to the expectation value  $\langle \bar{a}_r  a_r\rangle$ remaining finite as dissipation goes to zero. The diagrams for $\langle \bar{a}_r a_r\rangle$ in the path integral approach that correspond to the lowest order collision integral are shown in Fig \ref{figKinetic}. 

The divergence of $\langle \bar{a} a\rangle$ as $i\epsilon$ goes to zero was noted in \cite{Gurarie.1}. And in effect, this formulation of the kinetic equation in terms of the divergent part of $\langle \bar{a}_r a_r\rangle$ was used in \cite{RS.2}. This will be discussed further in Sec \ref{Sec corrections higher twopoint}.

\subsection{Equations for higher-order cumulants}\label{Sec ehrenfest JJ}

There is no reason to stop with $G=\bar{a}_r a_r \equiv J_r$ in \eqref{ehrenfest}. More complicated forms of $G$ may lead to additional equations. In particular the connected expectation values $\langle J_r J_s \rangle_c$ will lead to a non-trivial equation at the same order as the ordinary kinetic equation associated to $\langle J_r\rangle$.

\begin{figure}
	\centering
	\includegraphics[width=0.4\textwidth]{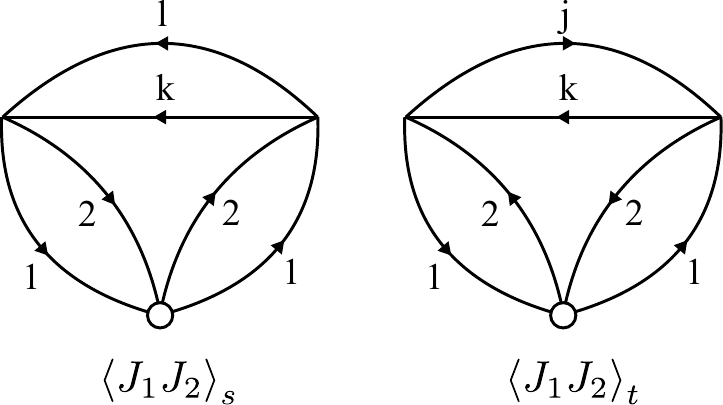}
	\caption{Connected diagrams for $\langle J_1 J_2\rangle^{(2)}$ in the four-wave theory. See \eqref{exp JJ}.}
	\label{figJJfourwave}
\end{figure}

To simplify the discussion, let us focus on the linear dissipation case, $\eta=0$, in which case the Ehrenfest equation for $G=J_1 J_2$ is,
\begin{align}
	\left\langle\{J_1 J_2,H\}\right\rangle&= 2\left(\gamma_1+\gamma_2\right)\left\langle J_1 J_2 \right\rangle-2\gamma_1n_1\left\langle J_2\right\rangle-2\gamma_2n_2\left\langle J_1\right\rangle\non&= 2\left(\gamma_1+\gamma_2\right)\left\langle J_1 J_2 \right\rangle_c.\label{J1J2c}
\end{align}
The second equality in terms of the connected expectation value is valid in the case that the collision integral \eqref{kinetic eq two sided} vanishes so that $\left\langle J_r\right\rangle = n_r$ to all orders in perturbation theory.

The calculation of the Poisson bracket on the left-hand-side proceeds similarly to the calculation of \eqref{kinetic eq three wave} and \eqref{kinetic eq four wave}. In the four-wave case it involves the six-point expectation values $\left\langle \bar{a}_2 a_2 \bar{a}_1 \bar{a}_j a_k a_l\right\rangle $, and similarly with $1$ and $2$ exchanged. At first order in $\lambda$ this expectation value involves the disconnected parts $\langle \bar{a}_j a_2\rangle^{(0)} \left\langle \bar{a}_1\bar{a}_2 a_k a_l\right\rangle^{(1)}$ and $2\langle \bar{a}_2 a_l\rangle^{(0)} \left\langle \bar{a}_1 \bar{a}_j  a_k a_2\right\rangle^{(1)}$, which respectively correspond to the ``s'' and ``t'' diagrams in Fig \ref{figJJfourwave} (see also Fig \ref{figFourPoint} later).

Using the result \eqref{four point exp} for the four-point expectation value, we may calculate the `collision integrals' associated to $J_1J_2$,
\begin{align}
	\left\langle\{J_1 J_2,H\}\right\rangle_s&= 16\,\text{Im}\left[ \sum_{kl}\left|\lambda_{12;kl}\right|^2\frac{\left(\frac{1}{n_1}+\frac{1}{n_2}\right)\left(\frac{1}{n_k}+\frac{1}{n_l}-\frac{1}{n_1}-\frac{1}{n_2}\right)}{\omega_{12;kl}+i\gamma_{12kl}}n_1^2n_2^2n_kn_l \right],\non
	\left\langle\{J_1 J_2,H\}\right\rangle_t&= 32\,\text{Im}\left[ \sum_{jk}\left|\lambda_{1j;k2}\right|^2\frac{\left(\frac{1}{n_1}-\frac{1}{n_2}\right)\left(\frac{1}{n_k}+\frac{1}{n_2}-\frac{1}{n_1}-\frac{1}{n_j}\right)}{\omega_{1j;k2}+i\gamma_{12kl}}n_1^2n_jn_kn_2^2 \right].\label{exp JJ}
\end{align}
These are at the same order as the collision integral for the ordinary kinetic equation, and unless the $n_r$ are of a special form such that these new collision integrals vanish in the zero dissipation limit, the expectation values $\langle J_1 J_2\rangle^{(2)}$ must diverge in the same limit.  Such divergences as $\gamma\rightarrow 0$ will be discussed further in Sec \ref{Sec discussion}, and it will be argued that they correspond to secular behavior in a time-dependent version of the theory.

\begin{figure}
	\centering
	\includegraphics[width=0.6\textwidth]{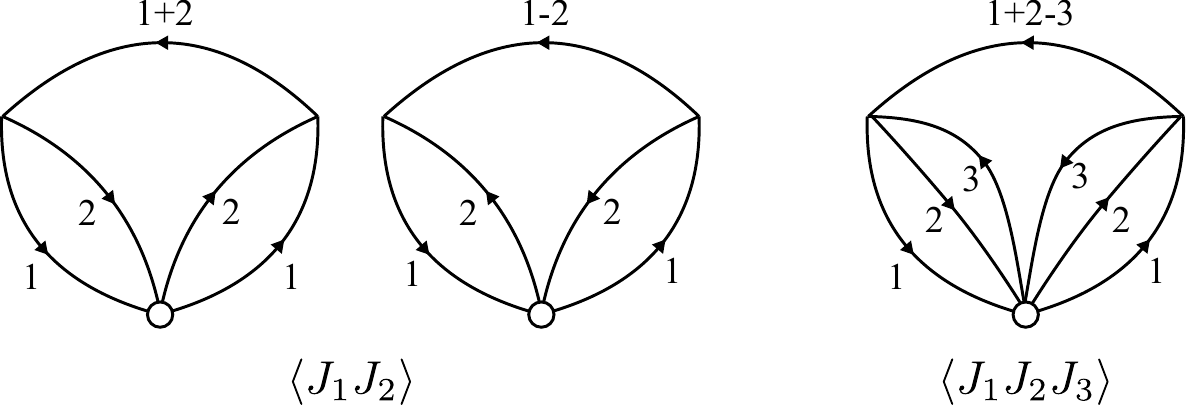}
	\caption{$\langle J_1 J_2\rangle_c^{(2)}$ in the three-wave theory and $\langle J_1 J_2 J_3\rangle_c^{(2)}$ in the four-wave theory.}
	\label{figJJJ}
\end{figure}

Finally note that the quantity $\langle J_1 J_2\rangle^{(2)}$ may also be considered in the three-wave theory, but in this case the connected amplitude will only diverge in the $\gamma\rightarrow 0$ limit if the external momentum $1$ and $2$ obey a resonance condition $\omega_1+\omega_2=\omega_{1+2}$ or $\omega_1-\omega_{2}=\omega_{1-2}$ (see Fig \ref{figJJJ}). A similar statement holds for the amplitude $\langle J_1J_2J_3\rangle^{(2)}_c$ in the four-wave theory. Such divergences on the resonance shell occur already in the first-order four- and three-point correlation functions \eqref{four point exp} and \eqref{three point exp}, and similar behavior has been noted in the wave turbulence literature using a very different approach \cite{ShavitFalkovich2020}.

\section{Corrections to the stationary state}\label{Sec corrections}
In this section we will explicitly calculate the stationary state distribution $\hat{H}\rho = 0$ to a fixed order in $\lambda$, and use it to calculate equal time expectation values. This is analogous to calculating the ground state wave-function in quantum mechanics perturbatively, and then calculating expectation values directly from the wave-function rather than using a path integral. All calculations done here will agree with the MSR approach of \cite{RSSS}, so the focus in this section is on introducing the method, and showing how the expression \eqref{kinetic eq two sided} relating the collision integral to the two-point expectation value is consistent. Some further calculations using this method are collected in Appendix \ref{Sec additional calculations}.

The perturbative expansion of the stationary state is
\begin{align}
	\rho &= \rho_0 - \hat{H}_0^{-1}\hat{V}\rho = \sum_{n=0}^{\infty}\left(-\hat{H}_0^{-1}\hat{V}\right)^n\rho_{0}.\label{LS}
\end{align}
This is a direct analogue to the Lippmann-Schwinger equation appearing in \cite{RS.2}, but we will see shortly that there will be no need to introduce an $i\epsilon$ term by hand.

To make sense of the $\hat{H}_0^{-1}$ operator in this equation we must first consider the eigenvalue problem associated to the free Fokker-Planck Hamiltonian. %This follows from a simple argument involving the probability conservation properties of the Fokker-Planck equation, and the uniqueness of the steady state $\hat{H}_0\rho_0=0$. The integral over phase space of $\hat{V}\rho$ must vanish since both the free and interacting Fokker-Planck equations conserve total probability.

%In order to consider the operator $\hat{H}_0^{-1}$ we will need to express $\hat{V}\rho$ in terms of eigenstates of $\hat{H}_0$ and invert the eigenvalue.

\subsection{Eigenstates of the free Fokker-Planck Hamiltonian}\label{Sec corrections eigenstates}

It will be convenient to express $a, \bar{a}$ in terms of action-angle variables $J,\alpha$,
\begin{align}
	a_k = \sqrt{J_k}e^{-i\alpha_k}.
\end{align}
The Poisson brackets in these variables are
\begin{align}
	\{F, G\}=\sum_k \der{F}{\alpha_k}\der{G}{J_k}-\der{G}{\alpha_k}\der{F}{J_k}.
\end{align}
We will further introduce a dimensionless action variable $x_k\equiv J_k/n_k$, in terms of which the stationary state \eqref{Def rho0} is just
\begin{align}
	\rho_0 = \prod_k e^{-x_k},\qquad x_k\equiv J_k/n_k.
\end{align}

Clearly the higher eigenstates of $\hat{H}_0$ will also just be a product over the mode index $k$, so let us momentarily focus on a single mode and supress the index. We wish to consider the time-independent Fokker-Planck equation,
\begin{align}
	\hat{H}_0\rho(x,\alpha)=\left[\omega\partial_\alpha -2\gamma\left(x\partial_x^2+(x+1)\partial_x +\frac{1}{4x}\partial_\alpha^2+1\right)\right]\rho(x,\alpha)=E_0\rho(x,\alpha).\label{time-independent FP}
\end{align}
This is solved by the following ansatz involving the integer $\nu$, and the undetermined function $\psi_{\kappa,\nu}$ and parameter $\kappa$,
\begin{gather}
\rho_{\kappa,\nu}(x,\alpha) = \sqrt{x}^{|\nu|}e^{i\nu\alpha}\psi_{\kappa,\nu}(x)e^{-x},\\
E_0=2\gamma\kappa+\gamma |\nu| + i\nu\omega.
\end{gather}
Now \eqref{time-independent FP} reduces to the associated Laguerre equation,
\begin{align*}
	x\psi_{\kappa,\nu}^{\prime\prime}+(1+|\nu|-x)\psi^\prime_{\kappa,\nu}+\kappa\psi_{\kappa,\nu}=0.
\end{align*}
So $\psi_{\kappa,\nu}(x)$ are taken to be associated Laguerre polynomials in $x$, and $\kappa$ is a non-negative integer. For the calculations in this paper the polynomials in Table \ref{Table Laguerre} will suffice.
\begin{table}
	\centering
	\begin{tabular}{c c c c}
$\kappa$ & $\nu$ & $\psi_{\kappa,\nu}(x)$ & $E_0$\\
\hline 
$0$ & $0$ & $1$ & $0$\\
$0$ & $\pm 1$ &  $1$ & $\gamma\pm i\omega$\\
$1$ & $0$ & $x-1$ & $2\gamma$\\
$1$ & $\pm 1$ & $x-2$ & $3\gamma\pm i\omega$\\
$2$ & $0$ & $x^2-4x+2$ & $4\gamma$
\end{tabular}
\caption{The first few associated Laguerre polynomials and the corresponding eigenvalues $E_0$ of $\hat{H}_0$.}\label{Table Laguerre}
\end{table}

\subsection{Collision integral from the four-point function}\label{Sec corrections 1st fourwave}
Now we can calculate the first order correction to the stationary state distribution using \eqref{LS},
\begin{align*}
	\rho = \rho_0 - \hat{H}_0^{-1}\hat{V}\rho_0+\order{\lambda^2}.
\end{align*}
We will begin with linear dissipation, where $\hat{V}=\hat{V}_L$,
\begin{align}
\hat{V}_L\rho_0 = \{\rho_0,V\} = \sum_r \frac{1}{n_r} \der{V}{\alpha_r}\rho_0.
\end{align}
In the four-wave theory \eqref{V four-wave}, this is
\begin{align}
	\hat{V}_L\rho_0 = i\sum_{ijkl}\lambda_{ij;kl}\left(\frac{1}{n_i}+\frac{1}{n_j}-\frac{1}{n_k}-\frac{1}{n_l}\right)\bar{a}_i\bar{a}_j a_k a_l \rho_0.\label{VL rho0}
\end{align}
Given the factor in parenthesis and the momentum conservation law implicit in $\lambda_{ij;kl}$ this will vanish if either $i$ or $j$ equals $k$ or $l$. Thus this is is in the form of a sum of eigenstates of $\hat{H}_0$, where modes $i,j,k,l$ have $\kappa=0$ and $\nu=\pm 1$. Using the sum of the eigenvalues $E_0= \gamma\pm i\omega$ for these modes, we can immediately write down the first order correction $\rho_1$,
\begin{align}
-\hat{H}_0^{-1}\hat{V}_L\rho_0=\sum_{ijkl}\frac{\lambda_{ij;kl}\left(\frac{1}{n_i}+\frac{1}{n_j}-\frac{1}{n_k}-\frac{1}{n_l}\right)}{\omega_{kl;ij}+i\gamma_{ijkl}}\bar{a}_i \bar{a}_j a_k a_l \rho_0,\label{rho1 4wave linear}
\end{align}
This uses the notation
$$\omega_{ij;kl}\equiv \omega_i+\omega_j-\omega_k-\omega_l,\qquad \gamma_{ijkl}\equiv \gamma_i+\gamma_j+\gamma_k+\gamma_l.$$

This is identical to Eq. (2.3) in Gurarie's work \cite{Gurarie.1}, but $\epsilon$ is identified with $\gamma_{ijkl}$ rather than being introduced ad hoc. Upon using $\rho_1$ to calculate the expectation value $\langle a_1 a_2 \bar{a}_3 \bar{a}_4\rangle$ we find identical results to the stochastic path integral approach \cite{RS.1},
\begin{align}
	\langle a_1 a_2 \bar{a}_3 \bar{a}_4\rangle^{(1)}=\frac{4\lambda_{12;34}\left(\frac{1}{n_1}+\frac{1}{n_2}-\frac{1}{n_3}-\frac{1}{n_4}\right)}{\omega_{34;12}+i\gamma_{1234}}n_1n_2n_3n_4.\label{four point exp}
	\end{align}
Using \eqref{kinetic eq four wave}, the lowest-order collision integral is then
\begin{align}
	\left	\langle \{J_1, H\}\right\rangle^{(2)}=4 \sum_{234}\text{Im}\left[\frac{4|\lambda_{12;34}|^2\left(\frac{1}{n_1}+\frac{1}{n_2}-\frac{1}{n_3}-\frac{1}{n_4}\right)}{\omega_{34;12}-i\gamma_{1234}}n_1n_2n_3n_4\right].\label{kinetic lowest order}
\end{align}

A similar lowest order calculation is done for the three-wave case in \ref{Sec corrections 1st threewave}, and the four-wave case with non-linear dissipation in Sec \ref{Sec corrections 1st diss}. Higher order corrections to the four-point function (and thus the collision integral) are calculated in Sec \ref{Sec corrections higher fourpoint}.
\subsection{Collision integral from the two-point function}\label{Sec corrections higher twopoint}

Now we wish to demonstrate that the same collision integral may be calculated from the two-point function using \eqref{kinetic eq two sided}. This requires that we calculate the second-order correction $\rho_2$.

In principle there is no obstacle to calculating $\rho$ to any finite order in $\lambda$ using \eqref{LS}, but calculating all terms in the correction quickly becomes tedious since, as seen concretely in the lowest-order three-wave case in Sec \ref{Sec corrections 1st threewave}, we must keep track of cases where indices in a multi-index summation become identical since they generically lead to different eigenstates of $\hat{H}_0$. This complexity is to be expected since a calculation of $\rho$ to a given perturbative order is equivalent to the calculation of all expectation values at that order. Instead of calculating everything at once it may be more efficient to calculate individual expectation values by focusing on particular eigenstates in the result.

Following \cite{RS.2}, note that the expectation value  $\langle \bar{a}a\rangle$ will only depend on the part of $\rho$ that is independent of the angles $\alpha$ in phase space. Let $\hat{P}_{\nu=0}$ be a projector on the eigenstates of $\hat{H}_0$ with $\nu=0$ for all modes. Using the previous result for $\rho_1$ \eqref{rho1 4wave linear}, the relevant terms are
\begin{align}
	\hat{P}_{\nu=0}\rho_{2}&= -\hat{H}_0^{-1}\hat{P}_{\nu=0}\hat{V}_L \rho_1\non
	&=-\hat{H}_0^{-1}\sum_{ijkl}\frac{4\left|\lambda_{ij;kl}\right|^2\left(\frac{1}{n_i}+\frac{1}{n_j}-\frac{1}{n_k}-\frac{1}{n_l}\right)}{\omega_{kl;ij}+i\gamma_{ijkl}}\{\bar{a}_i \bar{a}_j a_k a_l \,\rho_0, \,a_i a_j \bar{a}_k \bar{a}_l\}\non
	&=-\hat{H}_0^{-1}\sum_{ijkl}\frac{4\left|\lambda_{ij;kl}\right|^2\left(\frac{1}{n_i}+\frac{1}{n_j}-\frac{1}{n_k}-\frac{1}{n_l}\right)}{\omega_{kl;ij}+i\gamma_{ijkl}}i\left(\der{}{J_i}+\der{}{J_j}-\der{}{J_k}-\der{}{J_l}\right)J_i J_j J_k J_l \,\rho_0.
\end{align}
At this point in \cite{RS.2} it was argued that the $-\hat{H}_0^{-1}$ factor would lead to a divergence so it was dropped, and the remaining factors were set equal to zero. These were then multiplied by $J_r$ with some arbitrary index and integrated over phase space to lead to the kinetic equation.

From the Fokker-Planck perspective, there is finite dissipation so there is no divergence, and we may continue calculating the relevant terms in $\rho_2$. Since we will calculate the expectation of $J_r$ we specifically need the $\kappa=1$ terms.
\begin{align}
	\hat{P}_{\nu=0,\kappa=1}\rho_{2}&=\sum_{ijkl}\frac{4i\left|\lambda_{ij;kl}\right|^2\left(\frac{1}{n_i}+\frac{1}{n_j}-\frac{1}{n_k}-\frac{1}{n_l}\right)}{\omega_{kl;ij}+i\gamma_{ijkl}}\hat{P}_{\nu=0,\kappa=1}\hat{H}_0^{-1}\left((x_i-1)J_j J_k J_l+\dots\right)\,\rho_0\non
	&=\sum_{ijkl}\frac{4i\left|\lambda_{ij;kl}\right|^2\left(\frac{1}{n_i}+\frac{1}{n_j}-\frac{1}{n_k}-\frac{1}{n_l}\right)}{\omega_{kl;ij}+i\gamma_{ijkl}}n_in_jn_kn_l\left(\frac{x_i-1}{2\gamma_i n_i}+\frac{x_j-1}{2\gamma_j n_j}-\frac{x_k-1}{2\gamma_k n_k}-\frac{x_l-1}{2\gamma_l n_l}\right)\,\rho_0.
\end{align}
Using this distribution to calculate the expectation value of $\bar{a}_r a_r = n_r x_r$ and multiplying by $2\gamma_r$ as in \eqref{kinetic eq two sided}, we indeed find an expression agreeing with the collision integral \eqref{kinetic lowest order}.

\section{More on non-linear dissipation}\label{Sec non-linear diss}
Non-linear dissipation was introduced here in terms of the Fokker-Planck equation, but it may be used in the MSR approach as well. The extension of MSR approach to non-linear dissipation is mostly straightforward but there are two non-trivial points we make here. One is that the MSR path integral in general involves a Jacobian term which was correctly dropped in \cite{RS.1} for linear dissipation but which must be included in the non-linear case. The second point is that the diagrammatic evaluation rules of \cite{RSSS} may be easily extended to incorporate non-linear dissipation, and doing so gives a non-trivial check on the calculation which is not available in the linear dissipation case.
\subsection{Jacobian term in the path integral}\label{Sec non-linear diss jacobian}
The derivation of the stochastic path integral for wave turbulence is covered in more detail in \cite{RS.1}. In brief, the expectation value of a function  $G$  on phase space is calculated as
\begin{align}
	\langle G \rangle=\int \mathcal{D}a\mathcal{D}\bar{a}\der{\left(E,\,\bar{E}\right)}{\left(a,\,\bar{a}\right)}\left\langle \delta\left(E\right)\delta\left(\bar{E}\right)\right\rangle \, G(a,\bar{a}).\label{fadeev popov path integral}
\end{align}
The brackets on the right-hand side refer to an expectation value over the stochastic forcing function $f$. There are delta functions to enforce the equations of motion\footnote{In principle to solve this equation for a unique $a_k$ we need to specify some initial conditions. But after averaging over $f$ it is consistent to take the initial conditions in the distant past, which means the path integral is calculating in the stationary state.} \eqref{eq motion a nonpert},
\begin{align}
	E_{k}=\dot{a}_k+\left(i+\frac{\gamma_k}{\omega_k}\right)\bar{\partial}_k{H}- f_k=0.
\end{align}
Following \cite{RS.1} the expectation of the delta functions leads to the action,
\begin{gather}
\left\langle \delta\left(E\right)\delta\left(\bar{E}\right)\right\rangle\propto e^{-S},\non
S=\int dt \sum_k \frac{\left|E_k\right|_{f=0}^2}{F_k}.
\end{gather}
After Fourier transforming $E_k(z)=\int dt \,e^{izt}E_k(t)$, the action may be written
\begin{gather}
	S= \int \frac{dz}{2\pi}\sum_k \left(\bar{a}_k\frac{(z-\omega_k)^2+\gamma_k^2}{2\gamma_k n_k}a_k -i g_k a_k \partial_k V + i \bar{g}_k \bar{a}_k \bar{\partial}_k V + \frac{\omega_k^2+\eta{\gamma_k^2}}{2\gamma_k n_k \omega_k^2}\bar{\partial}_k V{\partial}_k V\right),\label{S non pert diss}\\
	g_k \equiv \frac{z-\omega_k+i\gamma_k}{2\gamma_k n_k i}\left(1+i\eta\frac{\gamma_k}{\omega_k}\right).\label{g non pert diss}
\end{gather}
This is written so as to hold for both linear and non-linear dissipation, which correspond to $\eta = 0$ and $\eta = 1$ respectively.

There is also a Jacobian factor in \eqref{fadeev popov path integral} since the integration is over the fields $a, \bar{a}$ rather than $E, \bar{E}$. It will indeed be valid to disregard this Jacobian for linear dissipation, but for non-linear dissipation it will lead to an additional term in the path integral. To calculate the Jacobian, as usual in the Faddeev-Popov procedure, it is convenient to introduce fermionic ghost fields. The action for the ghost fields $c, \bar{c}, d, \bar{d}$ is
\begin{align}
	S_{Jacobian}&= \int dt\left[\sum_{k} \bar{c}_k \left(\der{}{t} + i\omega_k +\gamma_k\right)  c_k +\bar{d}_k \left(\der{}{t} - i\omega_k +\gamma_k\right) d_k \right.\non
	 &\left.\qquad+ \sum_{kl}\bar{c}_k \left(i+\frac{\gamma_k}{\omega_k}\right) \left(\partial_l\bar{\partial}_k V c_l+\bar{\partial}_l\bar{\partial}_k V d_l\right) +\bar{d}_k \left(-i+\frac{\gamma_k}{\omega_k}\right) \left(\partial_l{\partial}_k V c_l+\bar{\partial}_l{\partial}_k V d_l\right) \right].
\end{align}
Considering the terms on the first line which involve no $a, \bar{a}$ dependence, the $c$ and $d$ fields have propagators
\begin{align}
	\langle c_l(t)\bar{c}_k (0) \rangle^{(0)} = \delta_{kl}\theta(t)e^{-i\omega_k t}e^{-\gamma_k t}, \qquad \langle d_l(t)\bar{d}_k (0) \rangle^{(0)} = \delta_{kl}\theta(t)e^{+i\omega_k t}e^{-\gamma_k t}.
\end{align}
These propagators vanish for $t<0$. The usual regularization $\theta(0)=\frac{1}{2}$ is taken.

Upon integrating the fermionic fields out, in principle all diagrams involving a single connected loop of fermion fields contribute, but since the propagators vanish for $t<0$ only the single vertex diagrams which involve a single $t=0$ propagator will contribute. Thus, up to a field-independent constant, the effective action due to the Jacobian is
\begin{align}
	S_{Jacobian}&= -\int dt \sum_{kl}\frac{1}{2}\delta_{kl}\left[ \left(i+\frac{\gamma_k}{\omega_k}\right) \partial_l\bar{\partial}_k V + \left(-i+\frac{\gamma_k}{\omega_k}\right)\bar{\partial}_l{\partial}_k V \right]\non
	&=-\int dt \sum_k \frac{\gamma_k}{\omega_k}\bar{\partial}_k\partial_k V.\label{Jacobian term}
\end{align}
Note that there is an overall minus sign due to the fermion loop. This Jacobian term may also be found by deriving the stochastic path integral directly from the Fokker-Planck Hamiltonian (see section 34.6 in \cite{ZJ.text}).

\subsection{Reduction to the thermal equilibrium case}\label{Sec non-linear diss eq}

In \cite{RSSS}, rather than integrating over frequencies $z$, integration was carried out in the time domain, and simple rules related to those of \cite{BS} were found that simplify the calculation of multi-loop diagrams in the three- and four-wave theories. These rules will now be extended to the non-linear dissipation case, and the effect of taking the thermal equilibrium limit will be discussed.

The rules in \cite{RSSS} relied on the fact that the vertex factors $g_k(z)$ given in \eqref{g non pert diss} simplify when $z$ is evaluated at the poles of the corresponding propagator. Clearly $g_k(\omega_k-i\gamma_k)=0$ vanishes, and at the other pole,
\begin{align}
g_k\left(\omega_k+i\gamma_k\right)=\Gamma_k \equiv \frac{1}{n_k}\left(1+i\eta \frac{\gamma_k}{\omega_k}\right).\label{Def Gamma}
\end{align}
The quantity $\Gamma_k$ appears also from the Fokker-Planck perspective in Sec \ref{Sec corrections 1st diss}. In \cite{RSSS}, only linear dissipation ($\eta=0$) was discussed, but the rules given there also hold for non-linear dissipation if $1/n$ is simply replaced by $\Gamma$ or $\bar{\Gamma}$, depending on whether it arose from $g$ or $\bar{g}$ (this is easily reconstructed from the sign of the term).

The replacement of $1/n$ by $\Gamma$ and $\bar{\Gamma}$ for $\eta=1$ may seem like an unnecessary complication, especially if we are ultimately interested in the limit in which dissipation goes to zero. But this allows for a strong check on our calculations. In the thermal equilibrium special case it has been shown above that the stationary state of the Fokker-Planck equation with non-linear dissipation is $\rho_{B}=e^{-H/T},$ and this state does not depend on the values of $\gamma_k$ at all. This means that all $\gamma$ dependence must disappear from expectation values when the temperatures $T_k=T$ are set to be uniform, or equivalently if $n_k$ is set to be equal to $T/\omega_{k}$. Since the expressions for higher-order corrections to expectation values can be rather complicated functions of $\gamma, \omega,$ and $n$ this is an extremely powerful consistency check.

	\begin{figure}
	\centering
	\includegraphics[width=0.2\textwidth]{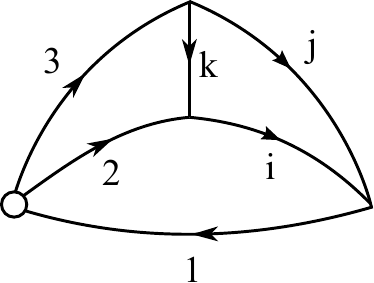}
	\caption{A ``tetrahedron'' diagram contributing to $\langle a_1 \bar{a}_2 \bar{a}_3\rangle^{(3)}$. Compare with Figure 3.b in \cite{RSSS}.}
	\label{figTetr}
\end{figure}

As an example, consider the correction to the three-point expectation value $\langle a \bar{a} \bar{a}\rangle_{tetr}$ corresponding to the diagram in Fig \ref{figTetr}. This was calculated for linear dissipation in Eq (3.9) of \cite{RSSS}. If we simply replace the appearances of $1/n$ in that result by $\Gamma$ and $\bar{\Gamma}$ according to sign, and also write $i\epsilon$ in terms of finite sums of $\gamma$, then we arrive at the correct result for finite non-linear dissipation,
\begin{align}
	\langle a_1 \bar{a}_2 \bar{a}_3\rangle_{tetr}=	&\sum_{ijk} \lambda_{1;ij}\lambda_{i;k2} \bar{\lambda}_{3;jk}    \frac{n_1n_2n_3n_in_jn_k}{\omega_{23;1}{+}i\gamma_{123}}\non&\qquad\times\left[ \frac{\bar{\Gamma}_1}{\omega_{23;ij}{+}i\gamma_{23ij}}\left( \frac{\left(\bar{\Gamma}_j+\bar{\Gamma}_k-\Gamma_3\right)\left(\bar{\Gamma}_i-\Gamma_2\right)}{\omega_{3;jk}{+}i\gamma_{3jk}} + \frac{\left(\bar{\Gamma}_i-\Gamma_2-\Gamma_k\right)\left(\bar{\Gamma}_j-\Gamma_3\right)}{\omega_{2k;i}{+}i\gamma_{2ki}} \right) \right.\non
	&\qquad- \frac{\Gamma_2}{\omega_{3i;1k}{+}i\gamma_{3i1k}}\left( \frac{\left(\bar{\Gamma}_j+\bar{\Gamma}_k-\Gamma_3\right)\left(\bar{\Gamma}_1-\Gamma_i\right)}{\omega_{3;jk}{+}i\gamma_{3jk}} + \frac{\left(\bar{\Gamma}_1-\Gamma_i-\Gamma_j\right)\left(\bar{\Gamma}_k-\Gamma_3\right)}{\omega_{ij;1}{+}i\gamma_{ij1}} \right) \non &\left.\qquad
	-\frac{\Gamma_3}{\omega_{2jk;1}{+}i\gamma_{2jk1}}\left( \frac{\left(\bar{\Gamma}_1-\Gamma_i-\Gamma_j\right)\left(-\Gamma_2-\Gamma_k\right)}{\omega_{ij;1}{+}i\gamma_{ij1}}  + \frac{\left(\bar{\Gamma}_i-\Gamma_2-\Gamma_k\right)\left(\bar{\Gamma}_1-\Gamma_j\right)}{\omega_{2k;i}{+}i\gamma_{2ki}} \right) \right].
\end{align}
The point of showing this convoluted expression is to impress upon the reader that upon substituting $n_r = \frac{T}{\omega_{r}}$ it simplifies dramatically and all $\gamma$ dependence cancels.
\begin{align}
	\langle a_1 \bar{a}_2 \bar{a}_3\rangle_{tetr}=	&\sum_{ijk} \lambda_{1;ij}\lambda_{i;k2} \bar{\lambda}_{3;jk}  n_1n_2n_3n_in_jn_k \left(-\frac{1}{T^3}\right).\label{three point simplification}
\end{align}

The result agrees with a calculation from the thermal equilibrium partition function,
\begin{align*}
\langle a_1 \bar{a}_2 \bar{a}_3\rangle= \frac{1}{Z}	\int \mathcal{D}a\mathcal{D}\bar{a} \left(a_1 \bar{a}_2 \bar{a}_3\right)e^{-\frac{1}{T}\left(H_0+V\right)},
\end{align*}
where a propagator with momentum $k$ is just $n_k$, hence the simplicity of \eqref{three point simplification}. All of the one-, two- and three-point expectation values in the three-wave theory calculated using the method of \cite{RSSS} have been tested  up to order $\lambda^3$  in this manner.

\section{Discussion}\label{Sec discussion}

In this paper we have discussed a Fokker-Planck Hamiltonian approach that is equivalent to the MSR path integral considered in \cite{RS.1,RSSS}, but more in the spirit of the phase space methods of \cite{Gurarie.1,RS.2}. The equivalence to the path integral approach is clear on general grounds since both follow from the same Langevin equation, but it was shown explicitly in Sec \ref{Sec corrections} and the associated Appendix \ref{Sec additional calculations} that an old-fashioned perturbation theory calculation using the Fokker-Planck Hamiltonian produces identical correlation functions to those calculated with the method of \cite{RS.1,RSSS}.

The advantage of considering the Fokker-Planck Hamiltonian lies in the clarification of non-perturbative aspects. It is quickly apparent that the thermal equilibrium distribution $\rho_B$ is not actually a stationary state of an MSR approach using finite linear dissipation, and this led to the introduction of non-linear dissipation. But to what extent do the results of this approach depend on the choice of dissipation?

Another question is raised by the non-perturbative stochastic Ehrenfest theorem discussed in Sec \ref{Sec ehrenfest} which gives a clearer perspective on the fate of equal time expectation values in limit of zero dissipation. What do the divergences in the limit of vanishing dissipation mean? And more broadly, what is the relation between the stationary state method presented here, and the time-dependent methods which are perhaps more common in wave turbulence? We will discuss these questions further.   

\subsection{Non-linear dissipation and renormalization}
	 Since we are mostly interested in the limit of vanishing dissipation, it may seem that choice of linear versus non-linear dissipation is irrelevant and merely affects intermediate steps of the calculation. For many expectation values this is true. As discussed in Sec \ref{Sec non-linear diss eq} and also seen in practice in Sec \ref{Sec corrections 1st diss}, the choice of dissipation merely affects the quantities $\Gamma_k$ in the numerators of expectation values. For linear dissipation $\Gamma_k=1/n_k$, but for non-linear dissipation $\Gamma_k$ has an additional imaginary part $i\gamma_k/(n_k \omega_k)$ which vanishes as the dissipation vanishes. So as long as the real parts in the sums of $\Gamma_k$ in the numerator do not vanish, both forms of dissipation lead to the same result. The choice is merely a calculation preference. Linear dissipation leads to simpler numerators, and non-linear dissipation allows for the consistency check of going to the thermal equilibrium limit, as discussed in Sec \ref{Sec non-linear diss eq}.
	
	But sometimes the real parts in the sums of $\Gamma_k$ do vanish, and there are indeed expectation values which still depend on the form of dissipation used even in the limit of vanishing $\gamma$. Simple examples occur already at first order in $\lambda$ in the four-wave theory (see Sec \ref{Sec corrections 1st diss}). For linear dissipation the corrections to the expectation values of $\bar{a}_k a_k$ and $ \bar{a}_k a_k  \bar{a}_l a_l$ vanish at first order, but they are non-zero for non-linear dissipation even in the limit of vanishing $\gamma$. So which is the physically correct choice of dissipation? Some physical arguments in favor of the non-linear dissipation choice have already been given Sec \ref{Sec non-linear diss intro}. Note that the expectation values in question are also non-vanishing in thermal equilibrium, which ought to be a special case of a stationary state where all modes are driven with the same temperature $T_k=T$.
	
	But in practice the difference between the forms of dissipation may be difficult to determine in classical field theories due to necessity of renormalization. For instance the correction to the expectation of $\bar{a} a$ for non-linear dissipation in the four-wave theory is given by \eqref{tadpole four wave}. For the sake of example let us consider the thermal equilibrium limit and a $\lambda$ that is momentum-independent except for an overall momentum conservation Kronecker delta. After taking the mode index to be continuous we get the integral
	\begin{align*}
		\langle \bar{a}(k_q) a(k_r)\rangle^{(1)} = -4 \frac{\lambda T^2}{\omega(k_r)^2} \delta^d(k_q-k_r)\,\int \frac{d^d k_s}{(2\pi)^d}\frac{1}{\omega(k_s)}.
	\end{align*}
	If the dispersion relation has the form of a power law $\omega(k_s)\sim k_s^n$ then the integral over $\omega(k_s)$ is divergent in either the UV or IR. This means that in practice we need some regularization involving a counterterm that will be set so that the theory agrees with the physical value of $\langle \bar{a} a \rangle$. This counterterm washes out the effect of the first-order correction to $\langle \bar{a} a \rangle$, and it is not quite so straightforward to distinguish the forms of dissipation based on this alone.
	
	This discussion is not meant to imply that divergences are absent in the linear dissipation case. Recently the issue of divergences in the four-wave theory was considered in \cite{FR}. There is no doubt that in order to actually solve for corrections to the KZ state using the corrections to the kinetic equation found in \cite{RS.1,RSSS}, the problem of IR and UV divergences in the momentum integrals must first be considered.

\subsection{The Ehrenfest theorem and the zero dissipation limit}

Physically one would expect that a non-equilibrium stationary state requires some real forcing and dissipation in the IR and UV to establish a flux through the inertial range, and thus it is no surprise if higher order corrections require regularization at these scales. But a perhaps more serious issue is that of divergences which occur as the auxiliary forcing and dissipation vanish. This auxiliary $\gamma$ is used to set up the non-equilibrium stationary state in the MSR approach, and it is intended to be set to zero at the end of a calculation.

Recall that in our general paradigm, we fix parameters $T_k$ in the dissipative part of the Fokker-Planck Hamiltonian \eqref{Hgamma linear} which set the temperatures for modes in the lowest order stationary state $\rho_0$, and then this is continued to some non-perturbative $\rho$ that satisfies $\hat{H}\rho=0$. Much as in thermal equilibrium, $\rho$ could fail to exist if the canonical Hamiltonian is not bounded from below in phase space,\footnote{Indeed, the three-wave theory is not bounded from below, and implicitly there should be additional interaction terms in $V$ which are higher-order in $\lambda$ in order for the theory to make sense. Depending on its physical origin the four-wave theory may also be the truncation of a theory with higher-order terms.} but otherwise it seems that for every choice of the parameters $T_k$ there is a distinct non-equilibrium stationary state.

Since the canonical Hamiltonian $H$ is not integrable this abundance of stationary states is initially rather puzzling, but the resolution is that the states for most choices of $T_k$ do not have a well-defined $\gamma\rightarrow 0$ limit. This is seen clearly in the stochastic Ehrenfest equation \eqref{ehrenfest}, which for linear dissipation takes the form
\begin{align}
	\left\langle\{G,H\}\right\rangle&= \sum_k \gamma_k \left(\left\langle a_k{\partial}_k G+\bar{a}_k\bar{\partial}_k G \right\rangle-2n_k\left\langle\bar{\partial}_k \partial_k G\right\rangle\right).\label{ehrenfest discussion}
\end{align}
Here $G$ is an arbitrary function on phase space, and if the stationary state is to survive the limit of zero dissipation and be a stationary state of the Liouville Hamiltonian, then the expectation of the time derivative $\dot{G}=\{G,H\}$ on the left-hand side must vanish. For $G=\bar{a}_r a_r$ this condition is equivalent to the vanishing of the collision integral of kinetic equation \eqref{kinetic eq general}, but there is a similar consistency condition for every function $G$.

The kinetic equation collision integral certainly does not vanish for arbitrary choice of $T_k$, or equivalently $n_k$, but it is well-defined. However the right-hand side of \eqref{ehrenfest discussion} shows that any non-zero expectation of $\{G,H\}$ in the limit $\gamma_k\rightarrow 0$ must be associated to divergences of expectation values in the same limit. In particular the lowest order collision integral is associated to divergences in the diagrams of Fig \ref{figKinetic}. So this shows clearly how the stationary states are pathological for most choices of $T_k$ in the $\gamma\rightarrow 0$ limit.

However the KZ solution, or any other stationary solution to the collision integral, involves a particular choice of $T_k$ such that the divergences in the two-point functions vanish at lowest non-trivial order in perturbation theory. This hints at a non-perturbative definition of the KZ solution as a non-equilibrium stationary state which has a regular two-point function even in the limit of $\gamma\rightarrow 0$.

But this hint is at best the beginning of a more complete treatment of the KZ stationary state. Note that even if the $T_k$ can be chosen so as to make the usual collision integral (e.g. \ref{kinetic lowest order}) vanish, there are other `collision integrals' associated with other choices of $G$, and which correspond to divergences in higher-order cumulants. An example is given in \eqref{exp JJ}. For an equilibrium solution the numerators of these additional collision integrals are proportional to the sum of frequencies in the resonance delta function and they vanish. But for a non-equilibrium solution it seems impossible that all of these collision integrals might be tuned to zero for some non-trivial choice of $T_k$.

There are two ways out of this impasse. The first way would be more exciting but it is just a sketch of a future approach. This involves positing that the KZ solution really can be extended to a stationary state in the full non-perturbative theory. Certainly this will involve some external forcing and dissipation which is not set to zero at the end of the calculation and which is necessary to establish a flux and provide regularization. But it may also involve a more flexible form than \eqref{Def rho0} for the lowest order $\rho_0$, which currently involves decoupled Gaussian modes. The current form for $\rho_0$ allows for an arbitrary choice of the spectrum $T_k$, and this is enough to set the lowest order collision integral to zero, but we may need a more general ansatz to ensure that the whole family of collision integrals vanish for all $G$ and all orders in $\lambda$.

To be clear, the full stationary state $\rho$ which we have been discussing throughout is not Gaussian. But correlation functions are approximately Gaussian in the weak coupling regime, and perhaps even this is too rigid. The possibility that weak wave turbulence is non-Gaussian is hardly a new idea. There was experimental observation of intermittency in capillary wave turbulence in \cite{FalconEtAl2007}. On the theoretical side, in \cite{ShavitFalkovich2020} the authors noticed linear growth with time on the resonance shell of mutual information in three-wave turbulence. That calculation is suggestive of the $\gamma^{-1}$ divergence seen in the three-point function \eqref{three point exp} on the resonance shell in the approach of this paper, and the authors similarly suggest the possibility of a non-Gaussian stationary state.

A second approach towards resolving the divergences at zero dissipation simply involves a reinterpretation of the time-independent paradigm we have been using. In the full Fokker-Planck equation we may begin in some non-stationary state (for instance the Gaussian $\rho_0$) at some finite time $t_0$ in the past, and the state will evolve in a time-dependent way towards the stationary state $\rho$. So far we have been taking $t_0$ to be in the distant past, and thus we have been calculating with $\rho$ from the very outset. But if we interchange the order of limits and instead hold $t_0$ finite and set the dissipation to zero first, then the Gaussian $\rho_0$ just evolves according to the Liouville equation and there is no conceptual difficulty with the vanishing dissipation limit.

The $\gamma^{-1}$ divergences which appear in expectation values in the time-independent paradigm correspond precisely to secular terms that grow as $t-t_0$ in the more conventional paradigm with finite $t_0$ and vanishing $\gamma$. This is easily seen in a modification of the MSR approach in which the interaction $V$ is taken vanish for $t<t_0$. A general diagram for $\langle \bar{a}_k(t) a_k(t)\rangle$ (such as those in Fig \ref{figKinetic}) may be calculated with the method of \cite{RSSS}, and it will lead to an integral like
\begin{align}\langle \bar{a}_k(t) a_k(t)\rangle=n_k+\int^t_{t_0} ds e^{-2\gamma_k (t-s)}\left[I_k + \dots\right].\label{finite t0 finite gamma}\end{align}
Here $I_k$ will end up being the collision integral, and the terms in ellipsis involve exponentially small factors like $e^{\Gamma t_0}$ where $\Gamma$ is some positive sum of the dissipations $\gamma_i$ for various modes $i$.
If we keep $t_0$ finite and set $\gamma\rightarrow 0$,
$$\langle \bar{a}_k(t) a_k(t)\rangle=n_k+I_k(t-t_0)+\dots.$$
Now the kinetic equation follows by taking a $t$ derivative of both sides and a little bit of hand waving.

If instead we keep finite dissipation and set $t_0\rightarrow -\infty$, we get a $\gamma^{-1}$ divergence as in \eqref{kinetic eq two sided},
$$\langle \bar{a}_k(t) a_k(t)\rangle=n_k+\frac{I_k}{2\gamma_k}.$$
However the terms in ellipses vanish, and the presence of $\gamma$ in $I_k$ allows for an easy interpretation in terms of delta functions in the $\gamma\rightarrow 0$ limit. In the more usual time-dependent paradigm some extra arguments are necessary to drop the terms in ellipses and extract the collision integral. Of course these are standard textbook arguments for the lowest order case, but this becomes more obscure at higher order in perturbation theory.

Similarly the $\gamma^{-1}$ divergences in higher order cumulants such as $\langle J_1 J_2\rangle_c$ in Sec \ref{Sec ehrenfest JJ}, correspond in the time-dependent paradigm to secular growth. The inability to set the whole family of collision integrals to zero for all $G$ and all orders in $\lambda$ reflects that for any choice of $n_k$ the Gaussian $\rho_0$ is not actually a stationary state. This is well-known but the new feature of this approach is the ease with which we can calculate the secular growth in higher order cumulants. It is hoped that this and the non-perturbative features of the MSR approach which are clarified by the Fokker-Planck perspective in this paper shed some light on a future more flexible approach towards the KZ stationary state.

\section*{Acknowledgments} 
I would like to thank Enrique Pujals, Michal Shavit, Misha Smolkin,  and especially Vladimir Rosenhaus for useful discussions. This work  is supported in part by NSF grant PHY-2209116 and by the ITS through a Simons grant. 

\appendix

\section{Additional perturbative calculations}\label{Sec additional calculations}

\subsection{Non-linear dissipation} \label{Sec corrections 1st diss}
If the four-point function in the four-wave theory is calculated using non-linear dissipation, the operator $\hat{V}=\hat{V}_L+\hat{V}_\gamma$ has an additional correction,
\begin{align}
	\hat{V}_\gamma \rho_0 &= -\sum_r \frac{\gamma_r}{\omega_r}\bar\partial_r \left(\rho_0\,\partial_r V\right)+\text{c.c}= -\sum_r \frac{\gamma_r}{\omega_r}\left(2\bar{\partial}_r\partial_r V - \frac{1}{n_r}\left(a_r\partial_r V+\bar{a}_r\bar\partial_r V\right)\right)\rho_0.
\end{align}
This will combine with $\hat{V}_L\rho_0$ to lead to
\begin{align}
	\hat{V}\rho_0= -8\sum_{rs}\frac{\gamma_r}{\omega_r}\lambda_{rs;rs}\bar{a}_s a_s \rho_0+i\sum_{ijkl}\lambda_{ij;kl}\left(\bar{\Gamma}_i+\bar{\Gamma}_j-\Gamma_k-\Gamma_l\right)\bar{a}_i \bar{a}_j a_k a_l \rho_0,
\end{align} 
where $\Gamma_i \equiv \frac{1}{n_i}\left(1+i\frac{\gamma_i}{\omega_i}\right)$. This factor $\Gamma$ was shown to arise in a natural way from the stochastic path integral in Sec \ref{Sec non-linear diss eq}.

The term involving $\Gamma$ is very similar to \eqref{VL rho0}, but it is no longer true that the summand vanishes when  $i$ or $j$ equals $k$ or $l$. When the indices are equal the summand is in some $\nu=0$ state of $\hat{H}_0$. We may split off the part of the sum involving equal indices, rewrite all powers of $\bar{a} a$ in terms of Laguerre polynomials, and act with $\hat{H}_0^{-1}$. The result is,
\begin{align}
	-\hat{H}_0^{-1}\hat{V}\rho_0&=\sum_{ij\neq kl}\frac{\lambda_{ij;kl}\left(\bar{\Gamma}_i+\bar{\Gamma}_j-\Gamma_k-\Gamma_l\right)}{\omega_{kl;ij}+i\gamma_{ijkl}}\bar{a}_i \bar{a}_j a_k a_l \rho_0 - 4\sum_{rs} \frac{ n_s}{\omega_r}\lambda_{rs;rs}(x_r-1)\rho_0\non
	&\quad -\sum_r \frac{ n_r}{\omega_r}\lambda_{rr;rr}\left(x_r^2-4x_r+2\right)\rho_0 - 2\sum_{r\neq s} \frac{2\gamma_r}{\gamma_{rs}}\frac{ n_s}{\omega_r }\lambda_{rs;rs}(x_r-1)(x_s-1)\rho_0 \label{rho1 4wave nonpert}
\end{align}

Upon taking the $\gamma\rightarrow 0$ limit, the $ij\neq kl$ summation on the first line is equivalent to the result for linear dissipation \eqref{rho1 4wave linear} and it leads to the same kinetic equation. The remaining terms represent differences between the two forms of dissipation which survive the $\gamma\rightarrow 0$ limit. The second summation on the first line involving $(x_r-1)$ leads to a first-order correction to $\langle \bar{a}_k a_k\rangle$ which we have already calculated in \eqref{tadpole four wave}. The two summations on the second line lead to a connected first-order correction for four-point expectation values with repeating indices,
\begin{align}
	\left\langle\left(\bar{a}_k a_k\right)^2\right\rangle^{(1)}&=4n_k	\left\langle \bar{a}_k a_k \right\rangle^{(1)}-4\frac{n_k}{\omega_k}\lambda_{kk;kk}\non
	\left\langle \bar{a}_k a_k  \bar{a}_l a_l\right\rangle_{k\neq l}^{(1)}&=n_k	\left\langle \bar{a}_l a_l\right\rangle^{(1)} + 	n_l\left\langle \bar{a}_k a_k \right\rangle^{(1)}- 4\frac{\frac{ \gamma_k}{\omega_k }n_l+\frac{ \gamma_l}{\omega_l }n_k}{\gamma_k+\gamma_l}\lambda_{kl;kl}.\label{4point repeating index}
\end{align}

Finally, note that in the thermal equilibrium special case where we set $n \rightarrow T/\omega$,
\begin{align*}
	\bar{\Gamma}_i+\bar{\Gamma}_j-\Gamma_k-\Gamma_l \rightarrow -\frac{1}{T}\left(\omega_{kl;ij}+i\gamma_{ijkl}\right).
\end{align*}
So up to a constant term, $\rho_1$ reduces to $-\frac{1}{T}V\rho_0$,
\begin{align*}
	\rho_1=-\hat{H}_0^{-1}\hat{V}\rho_0&\rightarrow-\frac{1}{T}\left(\sum_{ijkl}\lambda_{ij;kl}\bar{a}_i \bar{a}_j a_k a_l -2\sum_{rs} \lambda_{rs;rs}n_r n_s\right)\rho_0.
\end{align*}
This is exactly what is expected from the exact solution $\rho = Z^{-1} e^{-\frac{1}{T}V}\rho_0 $, where $Z$ is a normalization factor chosen to ensure that the integral over phase space of $\rho$ is equal to that of $\rho_0$.
\subsection{Three-wave theory} \label{Sec corrections 1st threewave}
The first-order correction to the three-wave theory \eqref{V three-wave} proceeds in much the same way as in the four-wave theory, but there are some extra complications arising from the zero mode. With linear dissipation,
\begin{align}
	\hat{V}_L\rho_0 = \frac{1}{2}\sum_{jkl}i\lambda_{j;kl}\left(\frac{1}{n_j}-\frac{1}{n_k}-\frac{1}{n_l}\right) \bar{a}_j a_k a_l  \rho_0 + \text{c.c}.\label{VL rho0 3wave}
\end{align}
When $j\neq k,l$ (or equivalently when $k,l \neq 0$), this is just a product of $\kappa=0, \nu=\pm 1$ eigenstates for three distinct frequencies. But unlike the four-wave case, the summand does not vanish when $j=k,l$, and we must consider other eigenstates as well. The first order correction may be calculated by separating the parts of the sum involving $k$ or $l=0$, and writing in terms of associated Laguerre polynomials,
\begin{align}
	-\hat{H}_0^{-1}\hat{V}_L\rho_0&=\left(\frac{1}{2}\sum_{j,kl\neq 0}\frac{\lambda_{j;kl}\left(\frac{1}{n_j}-\frac{1}{n_k}-\frac{1}{n_l}\right)}{\omega_{kl;j}+i\gamma_{jkl}} \bar{a}_j a_k a_l \rho_0\right.\non
	&\qquad-\frac{1}{2}\frac{\lambda_{0;00}}{\omega_{0}+3i\gamma_{0}} a_0 \left(x_0-2\right)\rho_0- \sum_{r\neq 0} \frac{\lambda_{r;r0} n_r }{n_0(\omega_0+i\gamma_0+2i\gamma_r)}a_0\left(x_r-1\right)\rho_0\non
	& \qquad\left. - \sum_r \frac{\lambda_{r;r0} n_r }{n_0(\omega_0+i\gamma_0)}a_0\rho_0\right)+\text{c.c.}\label{rho1 3wave linear}
\end{align}
The first term and the two terms on the second line lead to an expression for the connected three-point expectation value, with or without repeating indices,
\begin{align}
	\langle \bar{a}_k \bar{a}_l a_j\rangle_{c}^{(1)}= \frac{\lambda_{j;kl}\left(\frac{1}{n_j}-\frac{1}{n_k}-\frac{1}{n_l}\right)}{\omega_{kl;j}+i\gamma_{jkl}} n_j n_k n_l.\label{three point exp}
\end{align}
This agrees with the expression in \cite{RSSS} and likewise may be used to derive the kinetic equation. The final term in \eqref{rho1 3wave linear} leads to the expression  \eqref{tadpole three wave} for the tadpole $\langle a_0\rangle$.

The calculation of $\rho_1$ in the three-wave theory taking into account non-linear dissipation is similar to what has already been discussed, and we will not reproduce the formula here. Much like the four-wave case \eqref{rho1 4wave nonpert} the final result involves the replacement of the $1/n$ terms by $\Gamma$ or $\bar{\Gamma}$ and the tadpole term is corrected to agree with $\eta=1$ in \eqref{tadpole three wave}. In the thermal equilibrium limit it also reduces to $-\frac{1}{T}V\rho_0$, and unlike the four-wave case there is no additional constant term since $Z$ is not corrected at first order in the three-wave theory.
%Those earlier works go on to derive the well-known leading order contribution to the four-wave kinetic equation, but here simply note that if the summand does not vanish on the resonances $\omega_{kl;ij}=0$, $\rho_1$ will diverge in the zero dissipation limit, so .

\subsection{Higher-order corrections to the collision integral}\label{Sec corrections higher fourpoint}
To find the subleading corrections to the four-wave collision integral we may calculate $\langle J_r \rangle^{(3)}$ using the third-order correction $\rho_3$, as was done in \cite{RS.2}. Alternatively, we may calculate the four-point expectation values $\langle a_1 a_2 \bar{a}_3\bar{a}_4 \rangle^{(2)}$ directly from $\rho_2$, and calculate the kinetic equation using \eqref{kinetic eq four wave}. We will briefly describe this calculation here.%how terms in $\rho_2$ are associated to the loop diagrams for these expectation values appearing in the stochastic path integral approach.

\begin{figure}
	\centering
	\includegraphics[width=0.8\textwidth]{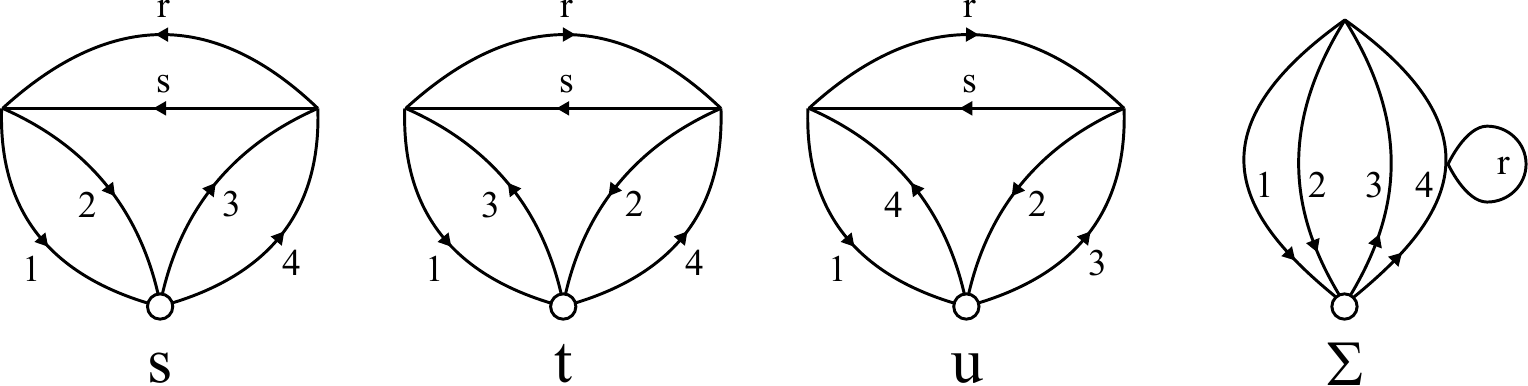}
	\caption{Diagrams for the corrections $\langle a_1 a_2 \bar{a}_3 \bar{a}_4\rangle^{(2)}$.}
	\label{figFourPoint}
\end{figure}
%The calculation of $\rho_2$ involves a summation over modes in $\rho_1$ \eqref{rho1 4wave linear} indexed by $ijkl$, and another summation over modes in $V$  which will be indexed by $pqrs$.

In order to calculate $\langle a_1 a_2\bar{a}_3\bar{a}_4 \rangle$ only the part of $\rho_2$ with $\kappa=0$ everywhere and $\nu=+1$ for modes $1,2$ and $\nu=-1$ for modes $3,4$ will contribute. So the relevant terms in $\rho_2$ will contain $\bar{a}_1 \bar{a}_2 a_3 a_4\rho_0$ and no additional $J$ dependence. The distinct ways that the indices $1,2,3,4$ can appear in the factors $\rho_1$ and $V$ in the expression $\rho_2=-\hat{H}_0^{-1}\{\rho_1,V\}$ will correspond to the distinct diagrams in the path integral approach in Fig. \ref{figFourPoint}.

To be specific, the total summation over indices in $V$ may be decomposed into a summation with four distinct indices, two distinct indices, and a single unique index.
\begin{align}
	V = \sum_{pq\neq rs} \lambda_{pq;rs} \bar{a}_p \bar{a}_q a_r a_s + 2\sum_{r\neq s} \lambda_{rs;rs} J_r J_s +\sum_r \lambda_{rr;rr} J_r^2. 
\end{align}
The part of $V$ involving repeated indices will correspond to the self-energy insertion diagrams referred to as ``$\Sigma$'' in Fig \ref{figFourPoint}, and the part with four distinct indices will lead to the $s,t,u$ diagrams. The specific possibilities for these terms in the Poisson bracket $\{\rho_1,\, V\}$ are,
\begin{itemize}
	\item $s$ diagram: $\{\bar{a}_1 \bar{a}_2 a_r a_s \,\rho_0,\, a_3  a_4  \bar{a}_r \bar{a}_s\},\, \{a_3  a_4  \bar{a}_r \bar{a}_s \,\rho_0,\, \bar{a}_1 \bar{a}_2 a_r a_s\}$
	\item $t$ diagram: $\{\bar{a}_1 a_3 \bar{a}_r a_s \,\rho_0,\, \bar{a}_2 a_4 a_r \bar{a}_s  \},\,\{\bar{a}_2 a_4 a_r \bar{a}_s \,\rho_0,\, \bar{a}_1 a_3 \bar{a}_r a_s  \}$
	\item $u$ diagram: $\{\bar{a}_1 a_4 \bar{a}_r a_s \,\rho_0,\, \bar{a}_2 a_3 a_r \bar{a}_s  \},\,\{\bar{a}_2 a_3 a_r \bar{a}_s \,\rho_0,\, \bar{a}_1 a_4 \bar{a}_r a_s  \}.$\label{V decomposed summation}
\end{itemize}	
Each diagram is associated to two distinct Poisson brackets which will end up corresponding to the two time-orderings of the internal vertices in the calculation scheme of \cite{RSSS}.

%\subsubsection{s-diagram}
As an example, consider the part of $\rho_2$ involving the first Poisson bracket in the $s$ diagram. Introduce a schematic projector $\hat{P}$ to project only on the relevant eigenstates of $\hat{H}_0$ and use the expression \eqref{rho1 4wave linear} for $\rho_1$,
\begin{align}
	\left(\hat{P}\rho_2\right)_{s}&=	-\hat{H}_0^{-1}\hat{P}\sum_{rs\neq 34}\frac{8\lambda_{12;rs}\lambda_{rs;34}\left(\frac{1}{n_1}+\frac{1}{n_2}-\frac{1}{n_r}-\frac{1}{n_s}\right)}{\omega_{rs;12}+i\gamma_{12rs}}\{\bar{a}_1 \bar{a}_2 a_r a_s \,\rho_0,\, a_3  a_4  \bar{a}_r \bar{a}_s\}\non & =-\hat{H}_0^{-1}\hat{P}\sum_{rs\neq 34}\frac{8\lambda_{12;rs}\lambda_{rs;34}\left(\frac{1}{n_1}+\frac{1}{n_2}-\frac{1}{n_r}-\frac{1}{n_s}\right)}{\omega_{rs;12}+i\gamma_{12rs}} i a_3  a_4  \bar{a}_r \bar{a}_s\left(\der{}{J_3}+\der{}{J_4}\right)\bar{a}_1 \bar{a}_2 a_r a_s \,\rho_0\non
	& =-i\hat{H}_0^{-1}\hat{P}\sum_{rs\neq 34}\frac{8\lambda_{12;rs}\lambda_{rs;34}\left(\frac{1}{n_1}+\frac{1}{n_2}-\frac{1}{n_r}-\frac{1}{n_s}\right)}{\omega_{rs;12}+i\gamma_{12rs}} \left(-\frac{1}{n_3}-\frac{1}{n_4}\right)J_r J_s \bar{a}_1 \bar{a}_2 a_3  a_4 \,\rho_0\non
	& =\sum_{rs\neq 34}\frac{8\lambda_{12;rs}\lambda_{rs;34}\left(\frac{1}{n_1}+\frac{1}{n_2}-\frac{1}{n_r}-\frac{1}{n_s}\right)}{\omega_{rs;12}+i\gamma_{12rs}} \frac{\left(-\frac{1}{n_3}-\frac{1}{n_4}\right)}{\omega_{34;12}+i\gamma_{1234}}\,n_r n_s \bar{a}_1 \bar{a}_2 a_3  a_4 \,\rho_0.\label{rho2 s calculation}
\end{align}
Upon taking the expectation value $\langle a_1 a_2 \bar{a}_3 \bar{a}_4\rangle$ we pick up factors $n_1 n_2 n_3 n_4$. This is an identical expression to what has been previously calculated using the stochastic path integral. See equation (3.5) in \cite{RSSS}. The $t$- and $u$-diagrams as well as the other Poisson bracket associated to the $s$-diagram may be calculated similarly.

The only remaining consideration is extending the summation to the special case where $r$ and $s$ equal $3$ and $4$, and analogous special cases for the other diagrams. These contributions are given by Poisson brackets involving indices only from the set $1,2,3,4.$ For instance: $\{\bar{a}_1 \bar{a}_2 a_3 a_4 \,\rho_0,\, a_3  a_4  \bar{a}_3 \bar{a}_4\}$. These special cases will also contribute to the self-energy insertion corrections. To tie up these loose ends, and because the formula has not appeared in the literature explicitly before, we will calculate the correction to the four-point expectation values due to self-energy (``$\Sigma$'') insertions in full.

Focusing on the $r\neq s$ summation in the expression \eqref{V decomposed summation} for $V$, the relevant corrections to $\rho_2$ are
\begin{align*}
	\left(\hat{P}\rho_2\right)_{\Sigma}&=	-\hat{H}_0^{-1}\hat{P}\sum_{r\neq s}\frac{8\lambda_{12;34}\lambda_{rs;rs}\left(\frac{1}{n_1}+\frac{1}{n_2}-\frac{1}{n_3}-\frac{1}{n_4}\right)}{\omega_{34;12}+i\gamma_{1234}}\{\bar{a}_1 \bar{a}_2 a_3 a_4 \,\rho_0,\, J_r J_s\}.
\end{align*}
The sum over modes in the Poisson bracket will vanish except for the $1,2,3,4$ indices. Let us focus on the $4$ index, corresponding to a self-energy insertion in the $4$ leg,
\begin{align}
	\left(\hat{P}\rho_2\right)_{\Sigma, 4}&=	-\hat{H}_0^{-1}\hat{P}\sum_{r\neq s}\frac{8\lambda_{12;34}\lambda_{rs;rs}\left(\frac{1}{n_1}+\frac{1}{n_2}-\frac{1}{n_3}-\frac{1}{n_4}\right)}{\omega_{34;12}+i\gamma_{1234}}\left(-i\,\bar{a}_1 \bar{a}_2 a_3 a_4 \,\rho_0 \der{}{J_4} J_r J_s\right)\non
	&=	-\hat{H}_0^{-1}\hat{P}\sum_{r\neq 4}\frac{16\lambda_{12;34}\lambda_{r4;r4}\left(\frac{1}{n_1}+\frac{1}{n_2}-\frac{1}{n_3}-\frac{1}{n_4}\right)}{\omega_{34;12}+i\gamma_{1234}}\left(-i\, J_r\bar{a}_1 \bar{a}_2 a_3 a_4 \,\rho_0\right)\non
	&=	-\sum_{r}\frac{16\lambda_{12;34}\lambda_{r4;r4}\left(\frac{1}{n_1}+\frac{1}{n_2}-\frac{1}{n_3}-\frac{1}{n_4}\right)}{\left(\omega_{34;12}+i\gamma_{1234}\right)^2}n_r\bar{a}_1 \bar{a}_2 a_3 a_4 \,\rho_0 + \dots \label{rho2 self-energy calculation}
\end{align}
Here the ellipsis refers to corrections in the $r=1,2,3,4$ cases that will contribute the missing terms in the $s,t,u$ diagrams, as mentioned above. Ignoring these details\footnote{The $r=1,2,3,4$ cases must be handled separately since the $\kappa=1, \nu=\pm1$ associated Laguerre polynomial is $x_r - 2$ rather than $x_r - 1$, leading to an extra factor of $2$.  The $r=4$ case is actually included in single $r$ summation in \eqref{V decomposed summation}, and the factor of $2$ is exactly what is needed to match the factor multiplying the $r\neq s$ terms. The excess in the $r=1,2,3$ cases will contribute the missing terms in the summation for the $s, t, u$ diagrams. For instance, the special cases where $r,s=3,4$ or $4,3$, which are missing in the summation in \eqref{rho2 s calculation}, are provided by the excess for the $r=3$ case in \eqref{rho2 self-energy calculation} and the corresponding excess for the $r=4$ case in the self-energy correction to leg $3$.}, the self-energy correction to the $4$ leg of the 4-point expectation value is
\begin{align}
	\langle a_1 a_2 \bar{a}_3\bar{a}_4 \rangle_{\Sigma, 4}=-\sum_{r}\frac{16\lambda_{12;34}\lambda_{r4;r4}\left(\frac{1}{n_1}+\frac{1}{n_2}-\frac{1}{n_3}-\frac{1}{n_4}\right)}{\left(\omega_{34;12}+i\gamma_{1234}\right)^2}n_rn_1 n_2 n_3 n_4.\label{4point self energy insertion}
\end{align}
Of course the correction to any other leg follows by relabeling the indices and possibly taking a complex conjugate. These results agree with a calculation using the method of \cite{RSSS}.

\section{Tadpole diagrams}\label{Sec non-perturbative tadpole}

\begin{figure}
	\centering
	\includegraphics[width=0.2\textwidth]{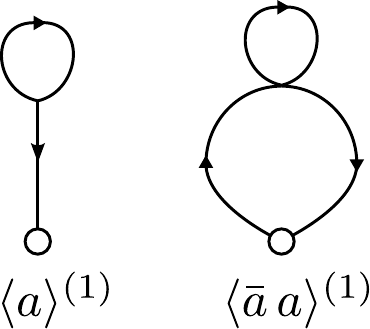}
	\caption{``Tadpole'' diagrams. Lowest order corrections to $\langle a\rangle$ in the three-wave theory \eqref{tadpole three wave} and $\langle \bar{a} a\rangle$ in the four-wave theory \eqref{tadpole four wave}.}
	\label{figTadpole}
\end{figure}

The calculations of the expectation values $\langle a\rangle$ in the three-wave theory and $\langle \bar{a}a\rangle$ in the four-wave theory have some pitfalls. In the path integral approach these expectation values involve the calculation of the \emph{tadpole} diagrams in Fig \ref{figTadpole} which involve a single propagator beginning and ending at the same vertex.

These expectation values will first be calculated from the Ehrenfest theorem and then from the MSR path integral approach. In the latter case it will be seen that the Jacobian term plays an essential role in the calculation in the non-linear dissipation case.

\begin{figure}
	\centering
	\includegraphics[width=0.3\textwidth]{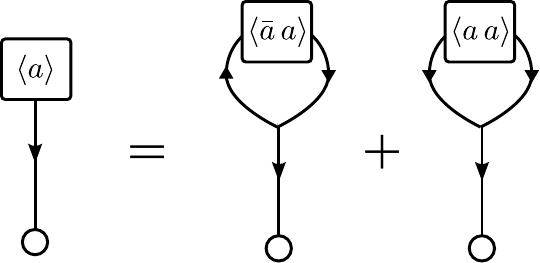}
	\caption{Equation \eqref{tadpole non perturbative} represented diagramatically. A box denotes the sum of all subdiagrams with the specified external legs leaving the box.}
	\label{figSchDyson}
\end{figure}

\subsection{Tadpoles from the Ehrenfest theorem}

As a simple application of the stochastic Ehrenfest theorem \eqref{ehrenfest}, take $G$ to be $a_k$ itself. After a brief calculation,
\begin{align}
	\left\langle\bar{\partial}_k H\right\rangle&=  \left(1-\eta\right)i\gamma_k\left\langle a_k\right\rangle.
\end{align}
This leads to a non-trivial expression for $\langle a_k \rangle$ in the three-wave theory \eqref{V three-wave}, 
\begin{align}
	\langle a_k\rangle =\frac{-1}{\omega_k-i\left(1-\eta\right)\gamma_k}\left(\sum_j \bar{\lambda}_{j;jk}\langle \bar{a}_j a_j\rangle+\frac{1}{2}\sum_{ij}\lambda_{k;ij}\langle a_i a_j\rangle \right).\label{tadpole non perturbative}
\end{align}
This equation could also have been derived in the diagrammatic approach of \cite{RSSS}, see Fig \ref{figSchDyson}.

In particular, at first order in $\lambda$, we find
\begin{align}
	\langle a_k\rangle^{(1)} =\frac{-1}{\omega_k-i\left(1-\eta\right)\gamma_k}\sum_j \bar{\lambda}_{j;jk}n_j.\label{tadpole three wave}
\end{align}

Note that as mentioned in \cite{RSSS}, the tadpole diagrams (or ``lollipops") in the three-wave theory may be canceled by the inclusion of a counterterm interaction in the Hamiltonian,
\begin{align*}
	V_{counterterm}=J_0 \bar{a} + \bar{J}_0 a.
\end{align*}
The results of this section imply that $J_0$ must be self-consistently tuned to the value
\begin{align}
	J_0=-\left(\sum_j \bar{\lambda}_{j;jk}\langle \bar{a}_j a_j\rangle+\frac{1}{2}\sum_{ij}\lambda_{k;ij}\langle a_i a_j\rangle \right).\label{tadpole counterterm}
\end{align}

The Ehrenfest theorem also leads to a simple calculation of the first-order correction to $\langle \bar{a}_r a_r\rangle $ in the four-wave theory. Since the left-hand side of \eqref{kinetic eq two sided} only contributes to order $\lambda^2$, we get immediately
\begin{align}
	\langle \bar{a}_r a_r\rangle^{(1)}=-\eta \frac{4n_r}{\omega_r}\sum_k\lambda_{kr;kr}n_k.\label{tadpole four wave}
\end{align}
For linear dissipation $\eta=0$, and there is no correction at all. This is consistent with a tadpole insertion only shifting the frequency of the propagator, which was mentioned in \cite{RS.1}. For non-linear dissipation $\eta=1$ and the result agrees with the thermal equilibrium special case.

\subsection{Tadpoles from the path integral Jacobian}

In the non-linear dissipation $\eta=1$ case, a path integral calculation for these expectation values will involve the Jacobian term \eqref{Jacobian term}.

In the three-wave theory the Jacobian term is
\begin{align}
	S_{Jacobian}= - \int dt \sum_k \frac{\gamma_k}{\omega_k}\left(\lambda_{k;k0} a_0 + \bar{\lambda}_{k;k0} \bar{a}_0\right).
\end{align}
This leads to a correction to $\langle a_0\rangle$,
\begin{align}
	\langle a_0\rangle_{Jacobian}=+\frac{2\gamma_0 n_0}{\omega_0^2+\gamma_0^2}\sum_k \bar{\lambda}_{k;k0}\frac{\gamma_k}{\omega_k}.
\end{align}
The full first order correction also involves the tadpole diagram on the left in Fig \ref{figTadpole}. Reading the propagator and vertex factors from the action \eqref{S non pert diss}, this evaluates to  
\begin{align*}
	\langle a_0 \rangle^{(1)}&=
	-i \sum_k \bar{\lambda}_{k;k0}\frac{2\gamma_0 n_0}{\omega_0^2+\gamma_0^2}\int \frac{dz}{2\pi}\frac{2\gamma_k n_k\left(\bar{g}_0+\bar{g}_k-g_k\right)}{(z-\omega_k)^2+\gamma_k^2}+\langle a_0\rangle_{Jacobian}.
\end{align*}
Using \eqref{g non pert diss}, the $\bar{g}_k-g_k$ terms in the numerator contain a logarithmically divergent term that is proportional to $z-\omega_k$, but this is taken to vanish due to a symmetric integration interval.\footnote{This step is equivalent to the regularization $\theta(0)=1/2$ used earlier.} This vanishing term is encountered also in the linear dissipation case, but the non-linear dissipation version of $\bar{g}_k-g_k$ contains an additional term that is exactly cancelled by $\langle a_0 \rangle_{Jacobian}.$

The remaining $\bar{g}_0$ term involves $z=0$. For non-linear dissipation,
\begin{align*}
	\bar{g}_0 = \frac{\omega_0^2+\gamma_0^2}{2\gamma_0 n_0 \omega_0 i},
\end{align*}
so in total,
\begin{align*}
	\langle a_0 \rangle^{(1)}&=
	- \sum_k \bar{\lambda}_{k;k0}\frac{n_k}{\omega_0},
\end{align*}
which agrees with \eqref{tadpole three wave}.

A similar calculation may be done in the four-wave theory for the first-order correction to $\langle \bar{a}_r a_r \rangle$. In this case the Jacobian term is quadratic,
\begin{align}
	S_{Jacobian}= - 4\int dt \sum_{kl}\frac{\gamma_k}{\omega_k}  \lambda_{kl;kl}\bar{a}_l a_l.
\end{align}

The insertion of a single propagator loop in a propagator with momentum index $r$, such as on the right of Fig \ref{figTadpole}, will lead to a self energy correction
\begin{align*}
	-4i\sum_s \lambda_{rs;rs}\int \frac{dz_s}{2\pi}\frac{2\gamma_s n_s\left(\bar{g}_r-g_r+\bar{g}_s-g_s\right)}{(z_s-\omega_s)^2+\gamma_s^2}.
\end{align*}
Once again the correction from the Jacobian term cancels the part of $\bar{g}_s-g_s$ that does not already cancel due to symmetric integration. So the total self-energy of order $\lambda$ is
\begin{align*}
	-4i\sum_s \lambda_{rs;rs}n_s\left(\bar{g}_r-g_r\right).
\end{align*}
This formula is true for both forms of dissipation. For linear dissipation, $\bar{g}_r-g_r$ is proportional to $z_s-\omega_s$ and $\langle\bar{a}_r a_r \rangle^{(1)}$ will vanish due to symmetric integration. For non-linear dissipation, there is a nonvanishing part of $\bar{g}_r-g_r$ left over, and
\begin{align}
	\langle \bar{a}_r a_r\rangle^{(1)}&=-4i\sum_s \lambda_{rs;rs}n_s\left(\frac{\gamma_r}{n_r\omega_r i}\right)\int\frac{dz_r}{2\pi}\left(\frac{2\gamma_r n_r}{(z_r-\omega_r)^2+\gamma_r^2}\right)^2\non
	&=-4\sum_s \lambda_{rs;rs}n_s\frac{n_r}{\omega_r},\nonumber
\end{align}
which agrees with \eqref{tadpole four wave}.

\end{document}